\documentclass[aps,pre,reprint,superscriptaddress]{revtex4-2}

\usepackage{graphicx}
\usepackage{subfigure}
\usepackage{caption}
\usepackage{color}
\usepackage[dvipsnames]{xcolor}
\usepackage{amsmath}
\usepackage{verbatim}
\usepackage{breakcites}
\usepackage{wrapfig}
\usepackage{soul}
\usepackage{xparse}
\usepackage{multirow}
\usepackage[shortlabels]{enumitem}

\newcommand{\fig}[1]{Fig.~\ref{#1}}

\newcommand{\eq}[1]{Eq.~\ref{#1}}

\newcommand{\movie}[1]{Movie~\ref{#1}}
\renewcommand{\vec}[1]{\underline{#1}}
\newcommand{\tens}[1]{\underline{\underline{#1}}}
\newcommand{\unity}{\tens{\hat{1}}}
\newcommand{\del}{\vec{\nabla}}

\newcommand{\kb}{k_\textmd{B}}
\newcommand{\kbt}{\kb T}

\newcommand{\pd}[2]{\partial_{#2} #1}

\newcommand{\av}[1]{\left\langle #1 \right\rangle}

\newcommand{\abs}[1]{\left| #1 \right|}

\newcommand{\nn}{\nonumber}
\newcommand{\ie}{\textit{i.e.} }
\newcommand{\eg}{\textit{e.g.} }

\newcommand{\dt}{\delta t}
\newcommand{\act}{\alpha}
\newcommand{\actMIN}{\act_\mathrm{eq}}
\newcommand{\actMin}{\act_\mathrm{turb}}
\newcommand{\actMax}{\act_\dagger}
\DeclareDocumentCommand\length{ g }{%
    {\ell%
        \IfNoValueF {#1} { _{#1} }%
    }%
}
\newcommand{\lenAct}{\length{\act}}
\newcommand{\lenSys}{\length{\text{sys}}}

\newcommand{\lenDef}{\length{d}}
\newcommand{\lenVel}{\length{v}}

\newcommand{\lenDir}{\length{n}}
\newcommand{\rcm}[1]{\vec{r}^{\textmd{cm}}_{#1}}
\newcommand{\vel}{v}
\newcommand{\velScale}[1]{\vel_{#1}}
\newcommand{\velAct}{\velScale{\act}}
\newcommand{\vcm}[1]{\vec{\vel}^{\textmd{cm}}_{#1}}

\newcommand{\velAv}{\vel_\mathrm{av}}
\newcommand{\vor}{\omega}

\DeclareDocumentCommand\corr{ m g }{%
    {C_{#1#1}%
        \IfNoValueF {#2} { \left(#2\right)}%
    }%
}
\newcommand{\NCell}{N_{c}}  
\newcommand{\NGM}{\chi_\mathrm{NGM}}

\newcommand{\tocite}[1]{\textcolor{cyan}{\textsuperscript{*}}}

\begin{document}


\title{Mesoscopic Simulations of Active-Nematics}

\author{Timofey Kozhukhov}
\affiliation{School of Physics and Astronomy, The University of Edinburgh, Peter Guthrie Tait Road, Edinburgh, EH9 3FD, UK.}

\author{Tyler N. Shendruk}
\affiliation{School of Physics and Astronomy, The University of Edinburgh, Peter Guthrie Tait Road, Edinburgh, EH9 3FD, UK.}

\date{\today}

\begin{abstract}
\noindent
Coarse-grained, mesoscale simulations are invaluable for studying soft condensed matter because of their ability to model systems in which a background solvent plays a significant role but is not the primary interest. 
Such methods generally model passive solvents; however, far-from-equilibrium systems may also be composed of complex solutes suspended in an active fluid. 
Yet, few coarse-grained simulation methods exist to model an active medium. 
We introduce an algorithm to simulate active nematics, which builds on multi-particle collision dynamics (MPCD) for passive fluctuating nematohydrodynamics by introducing dipolar activity in the local collision operator. 
Active-nematic MPCD (AN-MPCD) simulations exhibit the key characteristics of active nematic turbulence but, as a particle-based algorithm, also reproduce crucial attributes of active particle models. 
Thus, mesoscopic AN-MPCD is an approach that bridges microscopic and continuum descriptions, allowing novel simulations of composite active-passive systems.

\end{abstract}


\maketitle

\section{Introduction}\label{sctn:intro}

While fish shoals, bird flocks and insect swarms~\cite{Vicsek2012PhysicsReports, Ramaswamy2005AnnalsOfPhysics} are magnificent macroscopic examples of active systems that captivate onlookers with their collective behaviours, the majority of biophysical research focuses on active systems composed of microscopic agents. 
Microscopic, motile particles locally convert free energy from their surroundings into mechanical work~\cite{Marchetti2013RevModPhys-Hydrodynamics, Ramaswamy2010AnnRevCond, Bar2020AnnRev} and collective dynamics~\cite{Chate2008EPJB}. 
These active particles are subject to relatively large stochastic fluctuations that play a significant role in their dynamics.
Prominent examples of active stochastic systems include suspensions of swimming microbes~\cite{ElgetiGompper2015RepProgPhys}, bacterial colonies~\cite{Sengupta2018PhysRevX}, tissue monolayers~\cite{Garcia2015,atia2018,mongera2018}, and mixtures of cytoskeletal filaments and motor proteins~\cite{Dogic2012Nature-MicrotubuleActiveNematic}. 

In addition to possessing microscopic activity and stochastic dynamics, each of these examples displays orientational ordering~\cite{Sano2017PRE, Marenduzzo2018NatComm, Duclos2014}. 
While shape-anisotropy is not a strict prerequisite of active self-propulsion, an innate direction of self-propulsion typically is. 
Furthermore, directionality regularly materializes as apolar orientation, even when the microscopic agents possess polar motility~\cite{GinelliChate2010PRL}. 
Despite the fact that swimming bacteria~\cite{Marenduzzo2018NatComm}, kinesin motor proteins marching along microtubules~\cite{Dogic2012Nature-MicrotubuleActiveNematic} and most other constituent self-propelled particles~\cite{Bar2020AnnRev} possess distinctly polar behaviour, the hydrodynamic limit of suspensions of many such particles is nematic in nature. 
This is because the interactions --- including dipolar forces enacted on the surrounding fluid medium~\cite{Marchetti2012EPJB} --- are principally apolar~\cite{Chate2006PRL, Marchetti2012EPJB}. 
Thus, active nematics have proven a fruitful model for studying intrinsically out-of-equilibrium materials. 

Alongside biophysical experiments, numerical simulations of nematic systems have been essential in developing physical understanding of activity's consequences for living materials. 
Different studies have attempted to model the microscopic details to a greater or lesser extent. 
For example, simulations of simple self-propelled rods in the dry limit~\cite{GinelliChate2010PRL} have found many of the same properties as more detailed simulations of substrate-crawling bacilliforms~\cite{Shendruk2020SciRep-Twitcher}. 
Similarly, self-propelled rods in the wet limit~\cite{ShelleyBetterton2017PRF, JoostShendruk2016SoftMatter} agree with simulations of swimming bacteria~\cite{ZoettlYeomans2019NaturePhys, Gompper2015SoftMatter}. 
While the microscopic details of active agents can vary significantly often even simple models can reproduce the essential emergent behaviours~\cite{Bar2020AnnRev}. 

Simple microscopic models of stochastic, self-propelled particles (such as nematic variations on the Vicsek model~\cite{Vicsek1995PRL, Chate2006PRL, Chate2008EPJB, Chate2021PhysRevE}) have been particularly well-studied and can be coarse-grained into kinetic theories, which in turn lead to specific hydrodynamic equations of motion~\cite{Chate2019PRL}. 
On the other hand, symmetry considerations allow one to write generalised hydrodynamic equations for active fluids in the continuum limit~\cite{TonerTu1998PhysRevE, Marchetti2013RevModPhys-Hydrodynamics, Ramaswamy2005AnnalsOfPhysics, ValerianiMarenduzzo2011SoftMatter}. 
Unlike the microscopic models, these often omit stochastic effects within the fluid. 
Substantial work has explored the physics of bulk active fluids, such as steady-state creation and annihilation of topological defects~\cite{Giomi2014PhilTransactionsA,Thampi2014EPL, Selinger2018SoftMatter} and the emergence of low-Reynolds number \emph{active turbulence}~\cite{Yeomans2012PNAS, Thampi2014RoySocA, Frey2015PNAS, Alert2020NaturePhys, Alert2022AnnRevCondMat}. 
The emergence of complex spontaneous flows and topological singularities from field equations suggests that confining active nematics could produce complex and dynamic self-actualized dissipative structures. 
However, studies of confined active nematics have mostly been limited to simple, fixed geometries~\cite{Dogic2019PNAS, Giomi2018NaturePhys, Shendruk2017SoftMatter, Brady2016NatComm, Bartolo2021PNAS, DePablo2019PRX, Thijssen2021, Keogh2022} and more complicated microfluidic geometries have rarely been considered. 

Similarly, passive solutes suspended in active fluids have not received extensive consideration~\cite{ValerianiMarenduzzo2011SoftMatter, Loewe2021NJP}. 
Driven particles, such as colloids and disks within active nematics, have exhibited exciting characteristics, such as effective negative viscosity~\cite{MarenduzzoPRL2012} and higher order defects~\cite{Leheny2020SoftMatter}. 
Experimentally, particles embedded in dense suspensions of swimming bacteria have exhibited anomalous diffusion~\cite{WuLibchaber2001PRL, ValerianiMarenduzzo2011SoftMatter, Patteson2016SoftMatter, Peng2016PRL, Lagarde2020SoftMatter}. 
While polymers suspended in intrinsically out-of-equilibrium athermal baths have been studied\cite{Sakaue2017,Osmanovic2017,Anand2020,Shafiei2020,winkler2017}, hydrodynamic interactions mediated through the active medium are typically neglected~\cite{winkler2020,Gompper2022PRE}. 
This highlights the need for coarse-grained simulation techniques that are capable of simulating both active fluids in complex geometries and suspended solutes possessing complex shape or internal degrees of freedom. 

Here, we present a novel mesoscopic particle-based algorithm for stochastically simulating active, wet, compressible, nematic fluids. 


\begin{figure}[tb]
     \centering
     \begin{subfigure}
         \centering
         \includegraphics[width=0.95\linewidth]{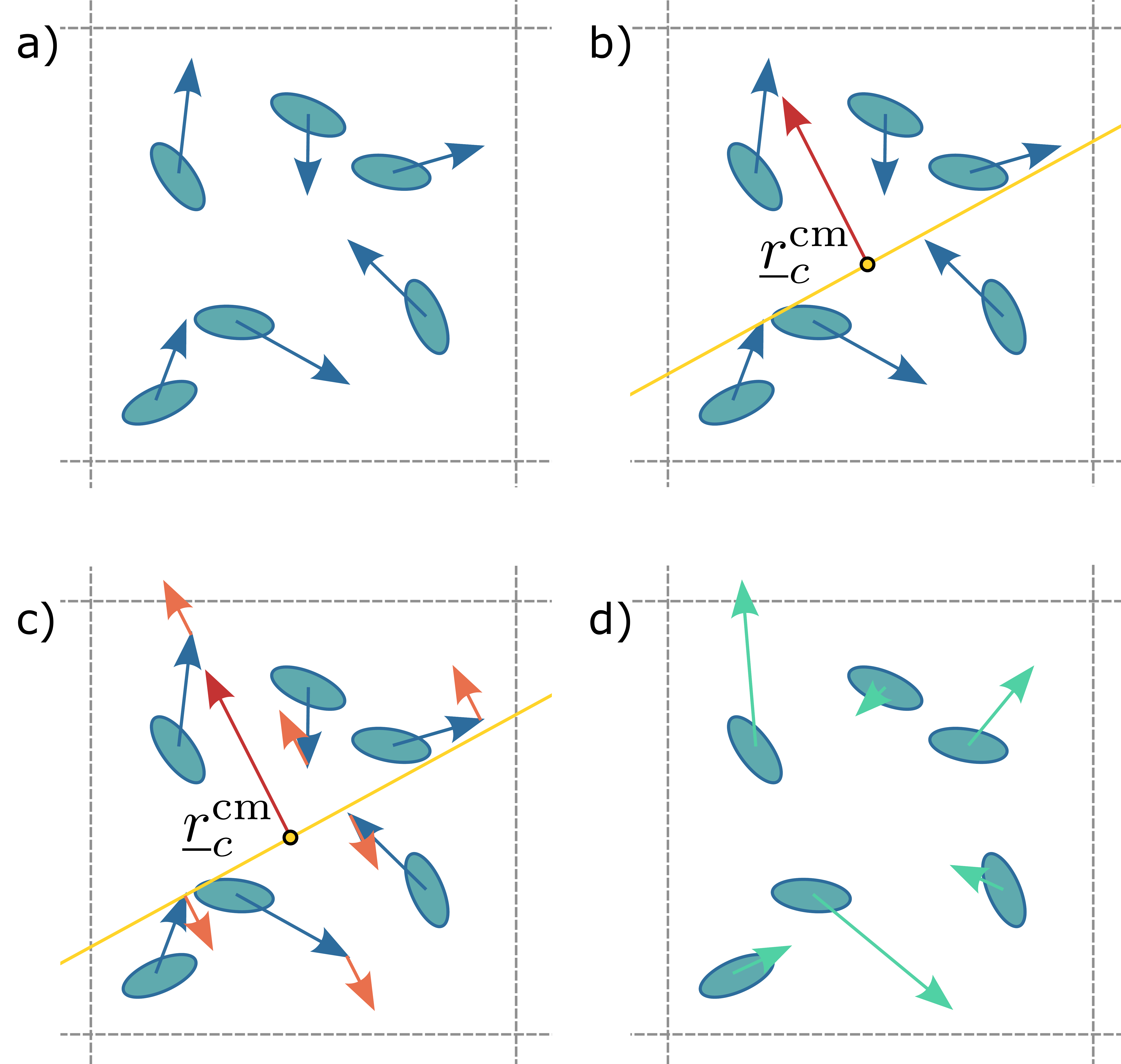}
         \label{fig:coll4}
     \end{subfigure}
     \caption{\textbf{Schematic representation of the active portion of the collision operator (\eq{eq:actOperator}).} 
     \textbf{(a)} Particles are binned into cells. Each particle has an internal orientation $\vec{u}_i$ shown schematically as an ellipsoidal shape and velocity $\vec{v}_i$ denoted by a blue arrow. 
     \textbf{(b)} The local nematic director $\vec{n}_c$ (red arrow) and centre of mass $\rcm{c}$ of the cell define a plane with approximately half the particles above and half below (yellow plane). 
     \textbf{(c)} Each particle is subject to an impulse away from the plane (orange arrows). 
     \textbf{(d)} Particles stream with their new velocities (green arrows). 
     }
    
     \label{fig:collOp}
\end{figure}

\section{Active Nematic MPCD Algorithm}\label{sctn:algorithm}

Continuum active nematohydrodynamics are described by three transport equations for mass, momentum and orientational order~\cite{Marchetti2013RevModPhys-Hydrodynamics}. 
The method introduced here to simulate these equations of motion extends the multi-particle collision dynamics (MPCD) algorithm to active nematohydrodynamics. 
The active collision operator locally injects energy, while conserving momentum. 
This section first introduces the conceptual basis of the passive MPCD framework, summarises the collision operators employed and finally extends these to an active-nematic collision operator.

\begin{figure*}[tb]
     \centering
     \begin{subfigure}
        \centering
        \includegraphics[width=0.98\textwidth]{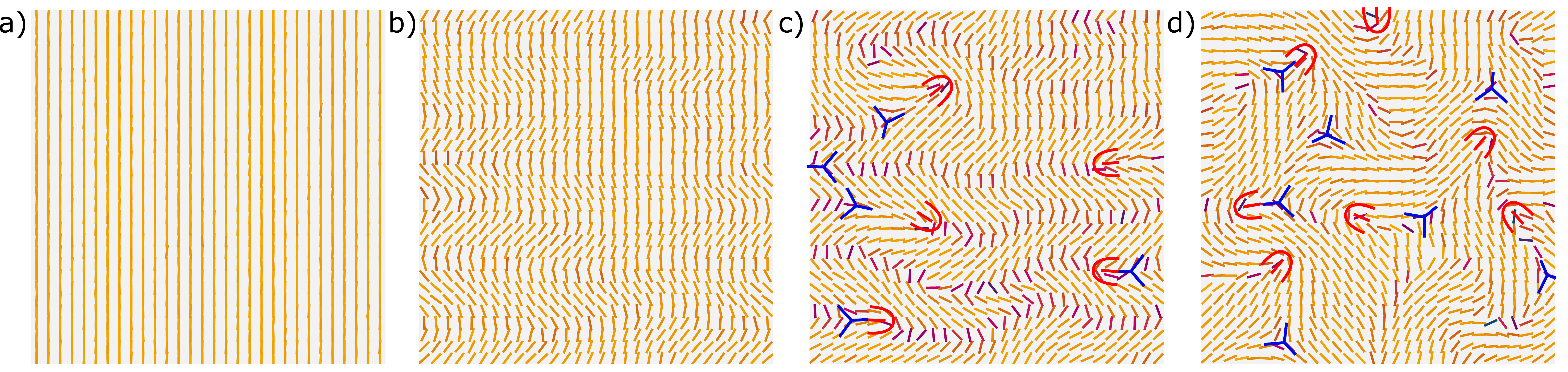}
     \end{subfigure}
     \caption{
        \textbf{Instabilities lead to defect pair creation and active turbulence.}
        Snapshots of the local director field $\vec{n}_c(\vec{r};t)$ coloured by scalar order parameter $S_c(\vec{r};t)$ for a system size of $\lenSys=30$ and extensile activity $\act=0.03$ from \movie{mov:defUnbindDir}.
        \textbf{(a)} \textbf{Timestep $0$}: The director is initialised in a fully aligned configuration and thermal noise perturbs the director field. 
        \textbf{(b)} \textbf{Timestep $110\dt$}: Extensile bend instability causes high-bend kink walls to form perpendicular to the global director with net force density parallel to the bend. 
        \textbf{(c)} \textbf{Timestep $160\dt$}: Defect pairs unbind within walls. $+1/2$ and $-1/2$ defects (red and blue respectively) are highlighted to guide the eye.
        \textbf{(d)} \textbf{Timestep $245\dt$}: Steady-state defect population arises, leading to active turbulence. 
    }
     \label{fig:pairCreation}
\end{figure*}

\subsection{Multi-particle collision dynamics}
Multi-particle collision dynamics algorithms discretise continuous hydrodynamic field into $N$ point-particles (labelled $i$). 
Each MPCD particle of mass $m_i$, position $\vec{r}_i$ and velocity $\vec{v}_i$ is said to stream ballistically for a time $\dt$ to a new position 
\begin{align}
    \vec{r}_i\left(t+\dt\right) = \vec{r}_i\left(t\right) + \vec{v}_i\left(t\right) \dt
\end{align}
before undergoing a multi-particle stochastic collision event~\cite{MalevanetsKapral1999JCP-MPCD}. 

Though real molecules that constitute a fluid interact with one another via microscopically-specific pair potentials, the details of these molecular interactions are commonly inconsequential to the resulting continuum equations of motion in the isotrop hydrodynamic limit. 
The MPCD method leverages this reality to recovered hydrodynamic equations on long-time and length-scales using an artificial and mesoscopic multiparticle collision operations, rather than specific intermolecular potentials between pairs of molecules~\cite{GompperIhle2009Book-MPCD, Kapral2008AdvChemPhys-MPCD}. 
MPCD collision events occur within lattice-based cells (labelled $c$) defined by a size $a$. 
In each cell, the instantaneous population $\NCell$ of particles stochastically exchange properties through collision operators that respect the relevant conservation laws. 
The collision operator governs the fluid transport coefficients~\cite{GompperIhle2009Book-MPCD}. 

To conserve mass and reproduced the continuity equation $\pd{\rho}{t} = -\del\cdot\left( \rho \vcm{c} \right)$, collision operators must simply leave the number of MPCD particles unchanged between collisions events. 
Like other particle-based hydrodynamic solvers, MPCD fluids are not strictly incompressible~\cite{Zantop2021JCP}. 

To conserve momentum, the cell's net momentum must be unchanged by collisions, which amounts to an unchanged centre of mass velocity $\vcm{c}=\av{ \vec{v} }_{c}$ if all the particles have the same mass $m_i=m$~\cite{MalevanetsKapral1999JCP-MPCD}. 
The average $\av{\cdot}_{c}$ is over the particles within cell $c$. 
The average velocity $\vcm{c}\left(\vec{r}_c;t\right)$ is interpreted as the hydrodynamic velocity field at the position of that cell $\vec{r}_{c}$. 
Similarly, constraints to the stochastic exchange of particle velocities lead to conservation of energy and angular momentum, and their associated hydrodynamic fields. 

\subsection{Angular-Momentum Conserving Andersen MPCD}
The continuous velocity field of an active nematic is given by a generalised Navier-Stokes equation in which the total hydrodynamic stress is the sum of a viscous stress, the orientational elastic stresses of nematic liquid crystals and an active stress~\cite{Marchetti2013RevModPhys-Hydrodynamics}. 
To simulate the stresses MPCD velocities are updated according to 
 \begin{align}
  \vec{v}_i\left(t+\dt\right) &= \vcm{c}\left(t\right) + \vec{\Xi}_{i,c}, 
  \label{eq:vel}
 \end{align}
where the collision operator $\vec{\Xi}_{i,c}$ is a non-physical exchange of momenta within each cell $c$. 
The collision operator is designed to be stochastic, while constrained to conserve net momentum (and kinetic energy, in passive fluids). 
The Andersen collision operator~\cite{Gompper2007EPL, Gompper2007PRE} is a common passive, isotropic collision operator with the form
\begin{align}
	\vec{\Xi}_{i,c} &= \vec{\xi}_i - \av{ \vec{\xi}_j }_{c} + \left( \tens{I}_c^{-1} \cdot \delta\vec{\mathcal{L}}_c \right)\times\vec{r}_i^\prime \nn\\
	                &\equiv \vec{\Xi}_{i,c}^0 . 
	\label{eq:passCol}
\end{align} 
Here, $\vec{\xi}_i$ is a random velocity drawn from the Maxwell-Boltzmann distribution for thermal energy $\kbt$ and $\av{ \vec{\xi}_j }_{c}$ is the cell average. 
The moment of inertia is $\tens{I}_{c}=\sum_j^{\NCell}m_j\left(r_j^{\prime2}\unity - \vec{r}^\prime_{j}\vec{r}^\prime_{j}\right)$ for point-particles in cell $c$ relative to their centre of mass $\rcm{c}$, where $\vec{r}_i^\prime=\vec{r}_i-\rcm{c}$. 
The third term in the collision operator corrects any spurious angular momentum $\delta \vec{\mathcal{L}}_c = \delta\vec{\mathcal{L}}_\text{vel}+\delta\vec{\mathcal{L}}_\text{ori}$ introduced by the collision operator $\delta\vec{\mathcal{L}}_\text{vel}=\sum_j^{\NCell}\vec{r}_j^\prime \times \left( \vec{v}_j - \vec{\xi}_j \right)$ or liquid crystalline backflow $\delta\vec{\mathcal{L}}_\text{ori}$ from the dynamics of the director field. 

The conceptual basis of MPCD can be extended from momentum collision operators that reproduce the hydrodynamic velocity field to collision operators for other hydrodynamic fields. 
In particular, nematic liquid crystals can be simulated via a collision operator to exchange particle orientations~\cite{Shendruk2015SoftMatter-NMPCD}. 

\begin{figure*}[tb]
    \centering
    \begin{subfigure}
        \centering
        \includegraphics[width=.95\textwidth]{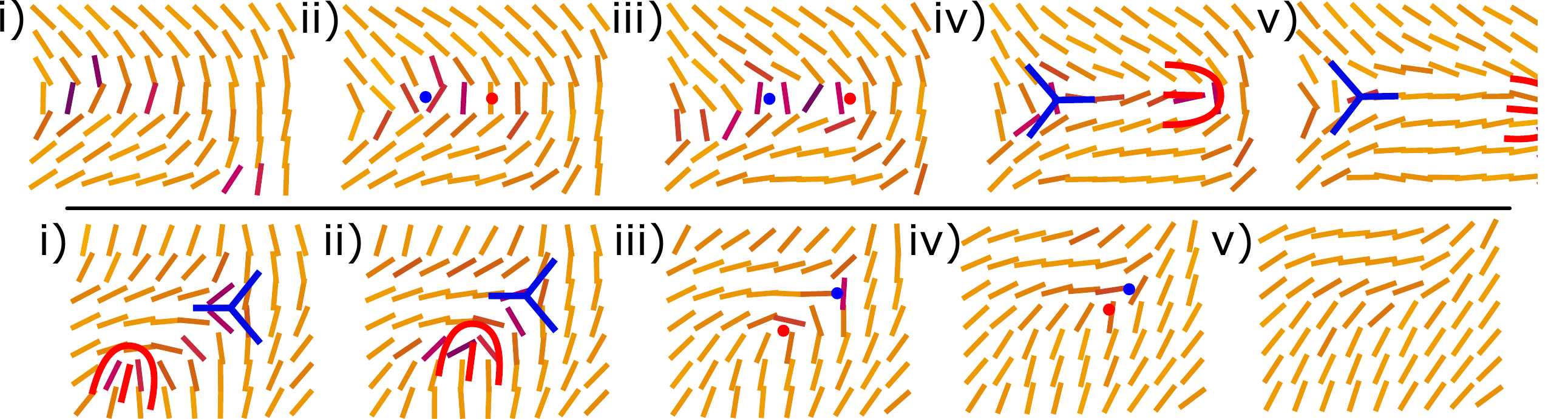}
    \end{subfigure}
    \caption{
       \textbf{AN-MPCD exhibits pair creation and annihilation events.}
       Snapshots of the local director field $\vec{n}_c(\vec{r};t)$ depicting creation (top row) and annihilation (bottom row) events in extensile AN-MPCD. 
       Snapshots are taken from the same simulation as shown in \fig{fig:pairCreation} and \movie{mov:defUnbindDir}. 
       The frame rate between snapshots is $5\dt$.
       Positive $+1/2$ (red) and negative $-1/2$ (blue) defects are highlighted to guide the eye.
       \textbf{(Top row):} 
       \textbf{(i)} An initial kink wall. 
       \textbf{(ii-iii)} The bend instability results in net force parallel to the bend, exacerbating the deformation. 
       \textbf{(iv)} Spontaneous creation of a $\pm 1/2$ defect pair. 
       \textbf{(v)} The $+1/2$ defect is self-motile and departs from the creation event site.
       \textbf{(Bottom row):}
       \textbf{(i)} Proximate $+1/2$ and $-1/2$ defects. 
       \textbf{(ii)} The self motile $+1/2$ defect approaches the $-1/2$ defect. 
       \textbf{(iii)} As the defects approach, they begin to lose their distinct shapes. 
       \textbf{(iv)} Defect annihilation reduces nematic distortion. 
       \textbf{(v)} Locally ordered nematic remains after an annihilation event. 
   }
    \label{fig:Creaton-Annihilation}
\end{figure*}

\subsection{Nematic MPCD}
In nematic liquid crystals, the tensor order parameter $\tens{Q}_c$ measures the extent and direction of orientational order through its largest eigenvalue $S_c$ and associated eigenvector $\vec{n}_c$. 
The evolution of $\tens{Q}_c\left(\vec{r}_c;t\right) = \av{ d \vec{u}_j\vec{u}_j - \unity}_{c}  /  \left(d-1\right)$ can be simulated using passive nematic MPCD by assigning an orientation $\vec{u}_i$ to each point-particle. 
The identity matrix is denoted $\unity$.
In this work we focus on an approach where the orientation is updated through a stochastic nematic multi-particle orientation collision operator~\cite{Shendruk2015SoftMatter-NMPCD} based on the local equilibrium distribution for the orientation field 
\begin{align}
    \label{eq:oriCol}
    \vec{u}_i\left(t+\dt\right) &= \vec{n}_c\left(t\right) + \vec{\eta}_{i}. 
\end{align}
The noise $\vec{\eta}_{i}$ is drawn from the Maier-Saupe distribution $\sim \exp{\left( \beta U S_{c} \left[\vec{u}_i\cdot\vec{n}_c\right]^2 \right)}$, where the width of the distribution about $\vec{n}_c\left(t\right)$ is controlled by a mean-field interaction potential, with an interaction constant $U$ and inverse thermal energy $\beta=1/\kbt$. 
The interaction constant $\beta U$ determines whether the fluid is in the isotropic or nematic phase, as well as the Frank elastic coefficient $K$~\cite{Shendruk2015SoftMatter-NMPCD, Hijar2020PhysicaA}. 
Nematic MPCD reproduces the one-constant approximation~\cite{Shendruk2015SoftMatter-NMPCD}. 

The orientations $\vec{u}_i$ are coupled to the velocity field via a bare tumbling parameter $\lambda$ and heuristic hydrodynamic susceptibility $\chi$. 
Backflow coupling is introduced through a viscous rotation coefficient $\gamma_\text{R}$~\cite{Shendruk2015SoftMatter-NMPCD}. 
The backflow coupling is mediated through angular momentum transfer from changes to orientation back into linear momentum through $\delta\vec{\mathcal{L}}_\text{ori}=\gamma_\text{R}\sum_j^{\NCell}\vec{u}_j \times \dot{\vec{u}}_j$.

Alternative hybrid finite difference/MPCD approaches have been proposed to apply MPCD to nematic liquid crystals~\cite{Marco2015JCP, Marco2019PRE}. 
Nematic MPCD has been employed to study nematohydrodynamic fluctuations and correlations~\cite{Hijar2019FluctNoiseLett, Hijar2019Arxiv}, electroconvection~\cite{Lee2017SoftMatter}, defects around nanocolloids~\cite{Hijar2020PhysicaA, Hijar2020PRE, Armendariz2021IntJModPhysB} and living liquid crystals~\cite{Mandal2021EurPhysJE}. 
We now extend the nematic Andersen-thermostatted collision operator to simulate wet active nematics. 

\subsection{Active Nematic MPCD}
To make MPCD active, an intrinsically out-of-equilibrium term must be included in the collision operation. 
We seek an algorithm for momentum-conserving (wet) active fluids~\cite{Marchetti2013RevModPhys-Hydrodynamics}; therefore, we propose a collision operator that injects energy but not momentum. 
The form of the introduced active stress should correspond to a force dipole density that can be extensile or contractile~\cite{Marchetti2013RevModPhys-Hydrodynamics}. 
As in continuum models of active nematics, this algorithm employs a form for which the local active stress is proportional to the nematic order tensor $\sim \tens{Q}_c$~\cite{Ramaswamy2003EPL} or, equivalently, a form for which the force dipole co-aligns with the local director $\vec{n}_c$. 

\begin{figure*}[tb]
     \centering
     \begin{subfigure}
         \centering
         \includegraphics[width=0.49\textwidth]{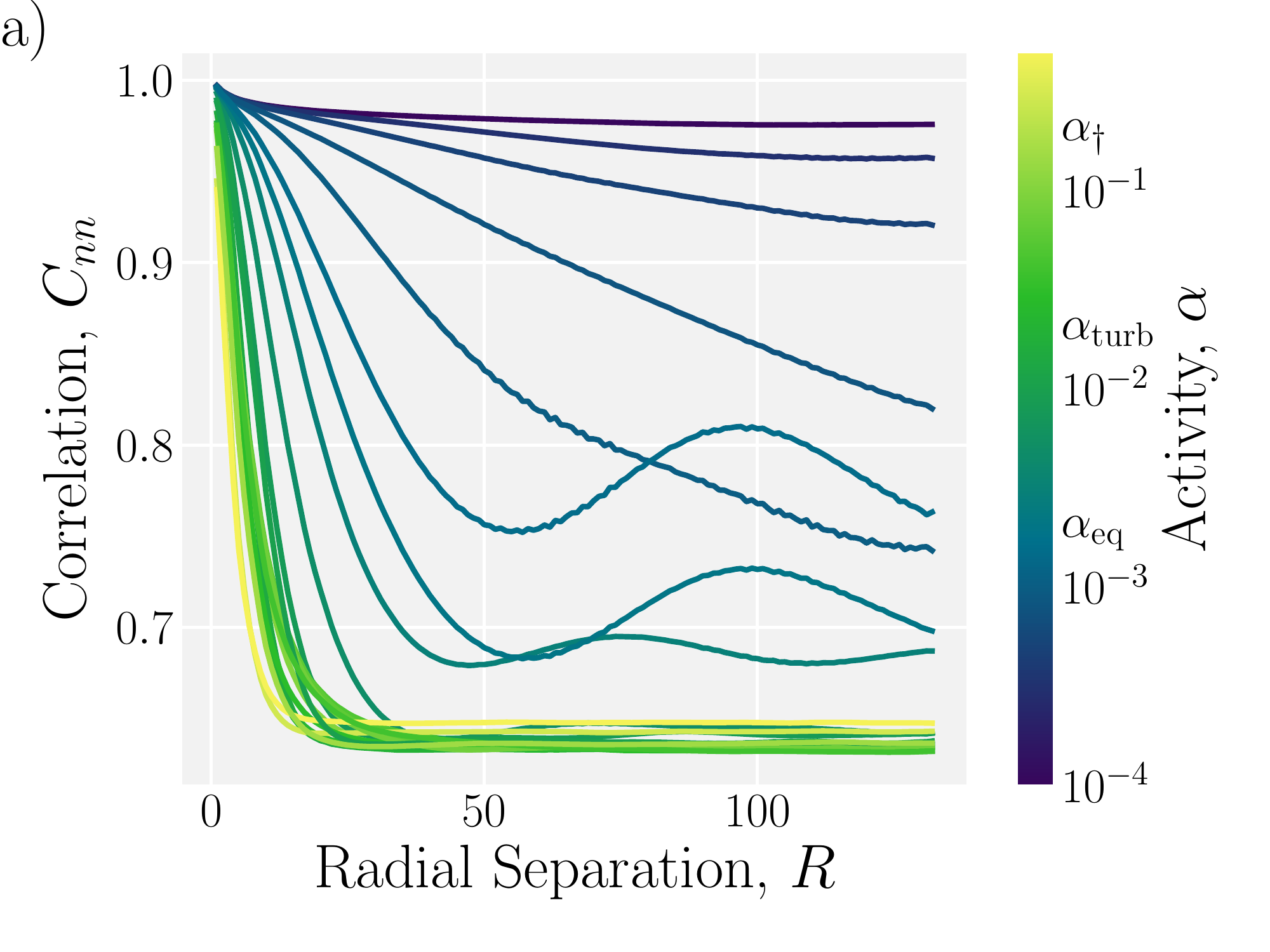}
         \label{fig:dirCor}
     \end{subfigure}
     \begin{subfigure}
         \centering
         \includegraphics[width=0.49\textwidth]{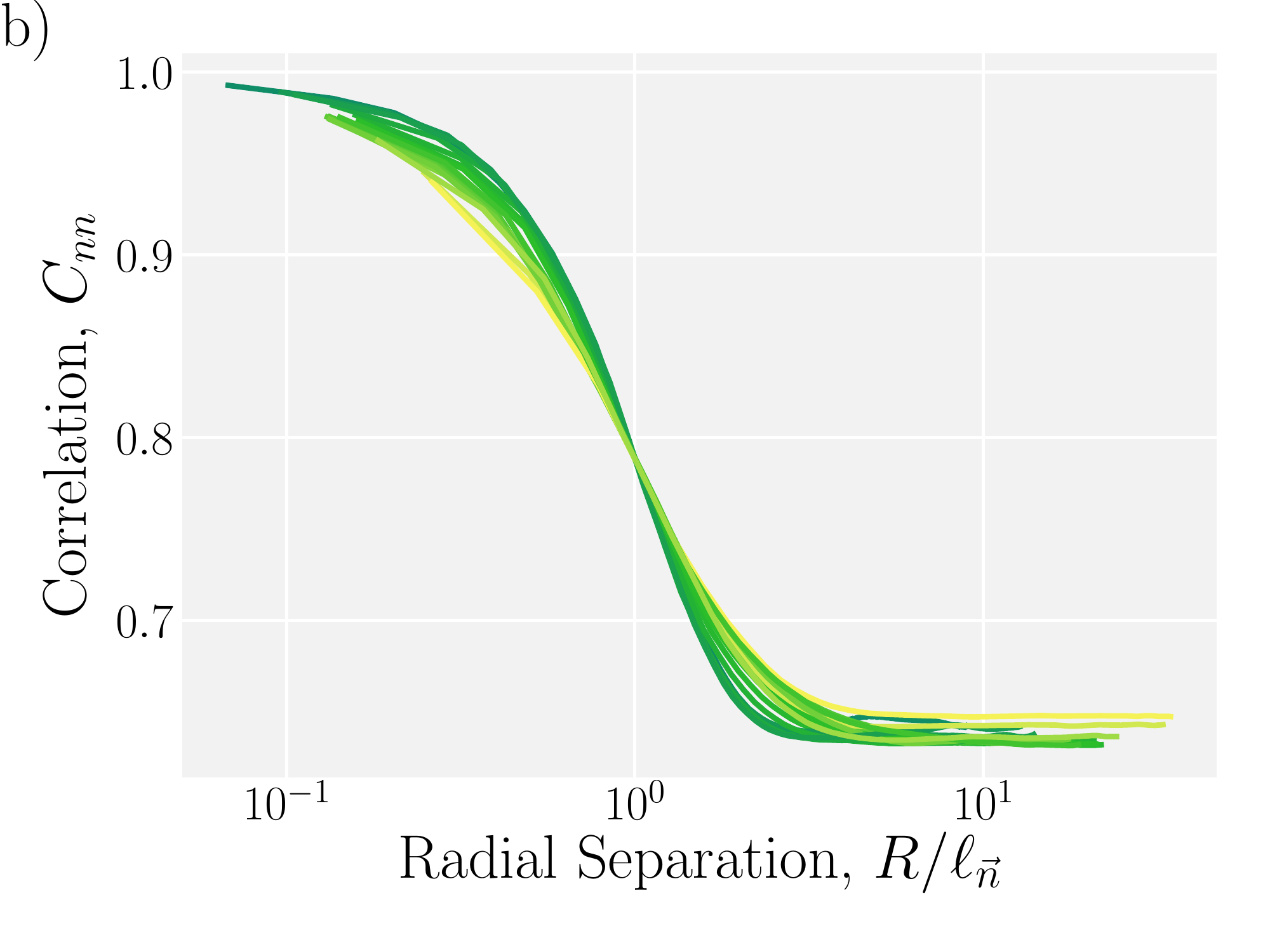}
     \end{subfigure}
     \caption{\textbf{Director spatial structure.} 
     \textbf{(a)} Director-director spatial correlation function $\corr{n}{R}$. 
     \textbf{(b)} Director-director spatial correlation function rescaled by nematic decorrelation length $\lenDir$. 
     }
     \label{fig:nemOrder}
\end{figure*}

To account for these considerations, the collision operator is a linear combination of passive and active contributions 
\begin{align}
	\vec{\Xi}_{i,c} &= \vec{\Xi}_{i,c}^0 + \vec{\Xi}_{i,c}^\text{A}. 
\end{align}
The passive contribution $\vec{\Xi}_{i,c}^0$ is given by \eq{eq:passCol}.
The  active nematic MPCD (AN-MPCD) collision operator 
\begin{align}
	\vec{\Xi}_{i,c}^\text{A} &= \act_c  \delta t \left( \frac{\kappa_i}{m_i} - \frac{\av{ \kappa_j }_{c}}{\av{m}_c} \right) \vec{n}_c 
	\label{eq:actOperator}
\end{align}
is composed of two terms: (i) individual impulses (per unit mass) $(\act_c \delta t / m_i) \kappa_i \vec{n}_c$ on each particle $i$ and (ii) a term to ensure local conservation of momentum $ -(\act_c \delta t / \av{m}_c) \av{ \kappa_j }_{c} \vec{n}_c$. 
Thus, as in the Andersen collision operator (\eq{eq:passCol}), any residual impulse is removed. 
The activity term $\left(\act_c \delta t/m_i\right) \ \kappa_i \vec{n}_c$ is composed of four factors (i) $\delta t$, (ii) $m_i^{-1}$, (iii) $\act_c $, and (iv) $\kappa_i\vec{n}_c = \pm\vec{n}_c$. 
\begin{enumerate}[(i)]
    \item The factor of $\delta t$ ensures activity $\act_c$ is given per unit MPCD time; rather than per unit iteration. 
    \item The factor of $m_i^{-1}$ gives $\act_c$ units of force. 
    \item The factor $\act_c(t)$ represents the local active dipole strength. 
    The cellular activity $\act_c$ is found by summing over the individual activity value $\act_j$ of each MPCD particle within a cell. 
    \begin{align}
		\act_c &= \sum_{j=0}^{\NCell} \act_j.
	    \label{eq:actCol}
    \end{align}
    This represents a local activity that is directly proportional to the local density of active agents within a given MPCD cell. 
    Positive particle activity $\act_j$ corresponds to extensile active nematics; whereas negative values correspond to contractile activity. 
    Here, all particles have the same activity $\act_j=\act$. 
    \item The factor $\kappa_i\left(\vec{r}_i,\rcm{c},\vec{n}_c\right)\vec{n}_c$ gives the direction of the active force acting on particle $i$. 
    In an active nematic, the activity is dipolar and is parallel to the local nematic director $\vec{n}_c(t)$. 
    Whether the impulse on each individual particle is parallel or antiparallel to the cell director is set through $\kappa_i\left(\vec{r}_i,\rcm{c},\vec{n}_c\right) = \pm1$. 
    This parallel/anti-parallel coefficient evenly splits the particles within cell $c$ into those that are driven ``forward'' ($\kappa_i=+1$) and those kicked ``backward'' ($\kappa_i=-1$).
    To do this, the collision operator considers the plane passing through the center of mass $\rcm{c}$ with surface normal $\vec{n}_c$. 
    Generally, half of the MPCD particles within cell $c$ are on either side of this plane. 
    The parallel/anti-parallel coefficient $\kappa_i=+1$ for particles above the plane and $\kappa_i=-1$ for particles below the plane. 
    This algorithm is visually illustrated in \fig{fig:collOp}. 
\end{enumerate}
Including the active collision operator (\eq{eq:actCol}) in the nematic MPCD algorithm results in a wet, particle-based, mesoscale, active nematic simulation method. 

\subsection{Simulation Units and Parameters}
Distances are reported in units of MPCD cell size $a \equiv 1$ and energy in units of thermal energy $\kbt\equiv1$. 
This work considers only a single species, such that all particles have the same mass $m_i = m \equiv 1$ and activity $\act_i=\act$ $\forall \ i$. 
Activity is given in MPCD units of force $ma/\tau^2 = \kbt / a \equiv 1$.
The MPCD time scale is thus $\tau=a\sqrt{m/\kbt}\equiv1$. 
Simulation parameters are expressed in these MPCD simulation units. 
Simulations are run in 2D in systems of size $\lenSys\times\lenSys$ with periodic boundary conditions and $\lenSys=200$, except when explicitly stated otherwise. 
Cartesian axes are denoted $\hat{x}$ and $\hat{y}$. 
The system-averaged fluid density is represented by the average particle density per cell, $\av{\NCell}=20$. 
The time step is set to $\dt=0.01$ and, unless otherwise stated, simulations utilize a warmup phase of $10^5$ time steps and a data production phase of $5\times10^5$ time steps. 
MPCD particles are initialized randomly within the control volume with random velocities, and orientations are initialised uniformly along the $y$-axis. 
The simulations are performed in the nematic phase by setting $U=10$, with a rotational friction $\gamma_\text{R}=0.01$, and hydrodynamic susceptibility $\chi=0.5$. 
The tumbling parameters $\lambda=2$ and $\lambda=1/2$ are simulated as representative of shear-aligning and flow-tumbling regimes respectively. 
The shear-aligning regime ($\lambda=2$) is assumed, except where explicitly stated. 
The activity $\abs{\act}<1$ is varied across four orders of magnitude to identify behavioural regimes of the algorithm. 
The activity is taken to be extensile ($\act>0$), unless otherwise stated. 
There exists a minimum activity $\actMIN$ below which the Andersen thermostat can absorb the active energy injection, yielding effective equilibrium behaviour. 
Active turbulence, with its associated steady-state population of defects, arises at some greater activity $\actMin\geq\actMIN$ and exists for a finite activity regime $\actMin\leq\act\leq\actMax$, where $\actMax$ denotes the end of the hydrodynamic active turbulence scaling regime. 

\section{Analysis Methods}\label{sctn:analysis}
\begin{figure}[tb]
    \centering
    \begin{subfigure}
        \centering
        \includegraphics[width=0.49\textwidth]{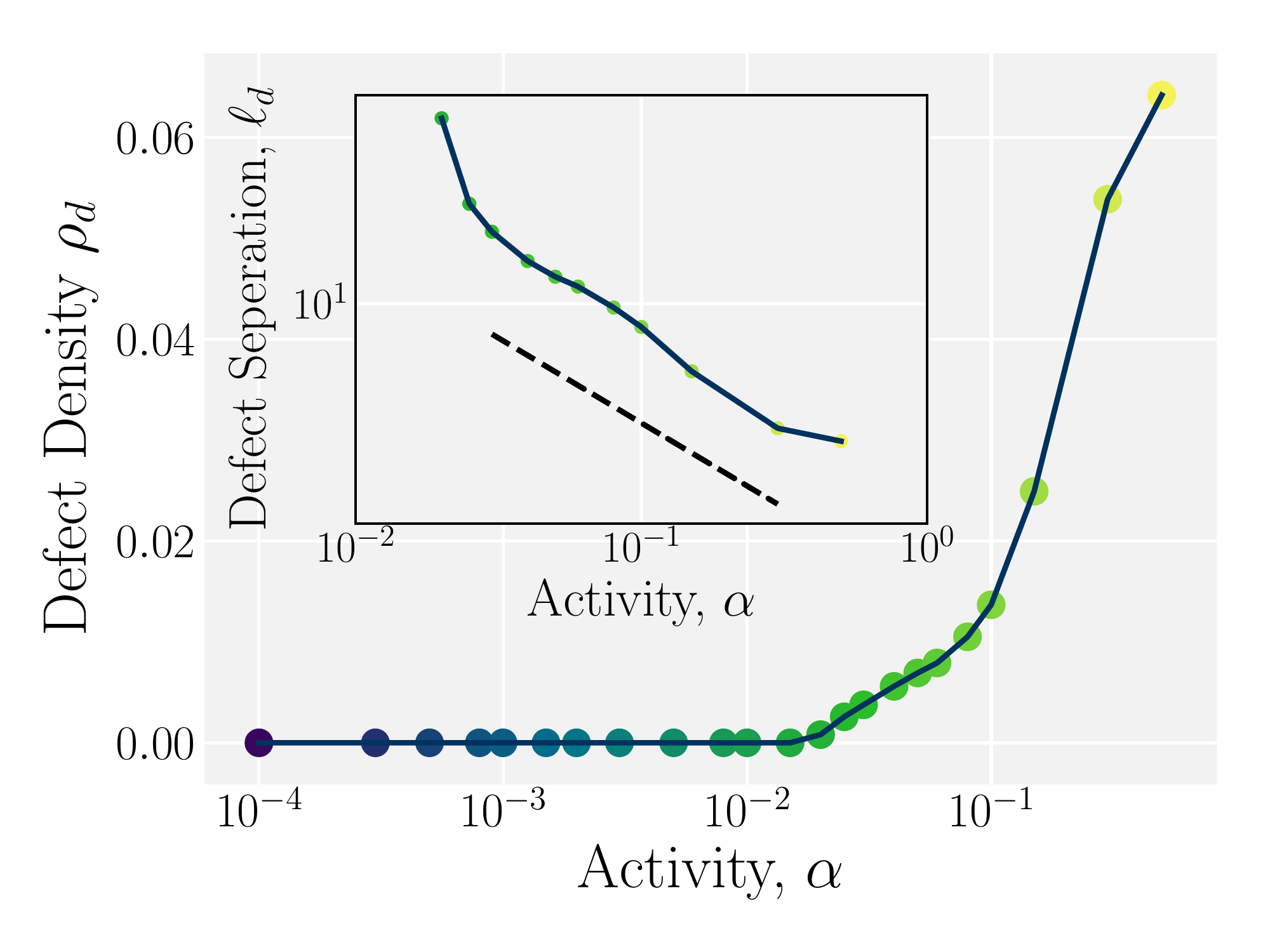}
        \label{fig:defDensSepInset}
    \end{subfigure}
    \caption{\textbf{Defect population and separation.} 
    Steady-state density $\rho_\text{d}$ of $\pm 1/2$ defects as a function of activity. 
    \textbf{(inset)} Root mean squared separation between all defects $\lenDef=\rho_\text{d}^{-1/2}$. 
    The dashed line indicates a scaling of $\lenDef\sim\alpha^{\mu}$ with $\mu=-1/2$.
    }
    \label{fig:defSep}
\end{figure}

\subsection{Characteristic Activity Scales}
Active nematic MPCD simulations are expected to possess a number of length scales. 
The algorithm itself has two; the system size $\lenSys$ and cell size $a\equiv1$. 
Additionally, an incompressible active nematic fluid has two: (i) the passive nematic persistence length which is proportional to the defect core size in the nematic phase; and (ii) the active length scale in fully developed mesoscale turbulence, which arises when the nematic elastic stress balances the active stress
\begin{align}
	\lenAct &\sim \sqrt{\frac{K}{\alpha}}. 
	\label{eq:actLength}
\end{align}
From this idealized dimensional analysis, the active length scales as $\lenAct\sim\alpha^{\mu}$ with the expected power law $\mu=-1/2$ in the regime of fully developed active turbulence.
However, it has been demonstrated that $\mu$ can saturate if activity is not sufficiently large and the nematic structure spans a substantial fraction of the system size~\cite{Hemingway2016SoftMatter}. 
Dimensional analysis also reveals an expected characteristic active velocity scale
\begin{align}
	\velAct &\sim \frac{\abs{\act} \lenAct}{\eta}, 
	\label{eq:actVelocity}
\end{align}
which arises when active stress is balanced by the viscous stress, characterized by the viscosity $\eta$. 
This is comparable to the speed of self-propelled $+1/2$ defects~\cite{Giomi2014PhilTransactionsA}. 
Since \eq{eq:actVelocity} includes both a direct factor of activity and an indirect factor through the active length scale (\eq{eq:actLength}), the velocity scales as $\velAct\sim\act^\gamma$, where ideally $\gamma=1/2$. 
A mesoscale algorithm for simulating the hydrodynamic limit of active nematics should be consistent with $\mu\approx-1/2$ and $\gamma\approx1/2$ in the regime of simulating fully developed turbulence (\ie $\actMin\leq\act\leq\actMax$). 

\begin{figure}[tb]
    \centering
    \begin{subfigure}
        \centering
        \includegraphics[width=0.49\textwidth]{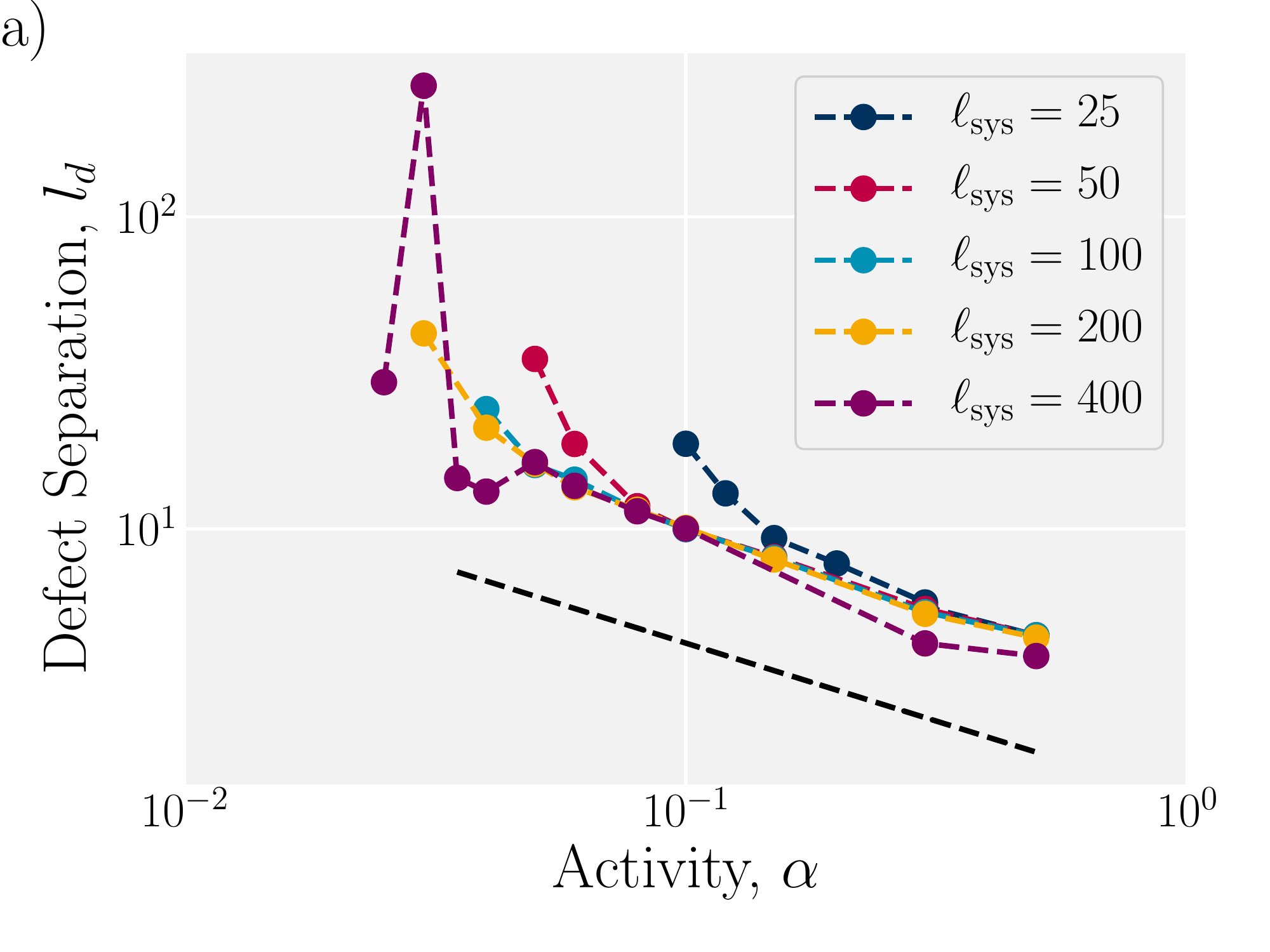}
        \label{fig:defSepTumble}
    \end{subfigure}
    \begin{subfigure}
        \centering
        \includegraphics[width=0.49\textwidth]{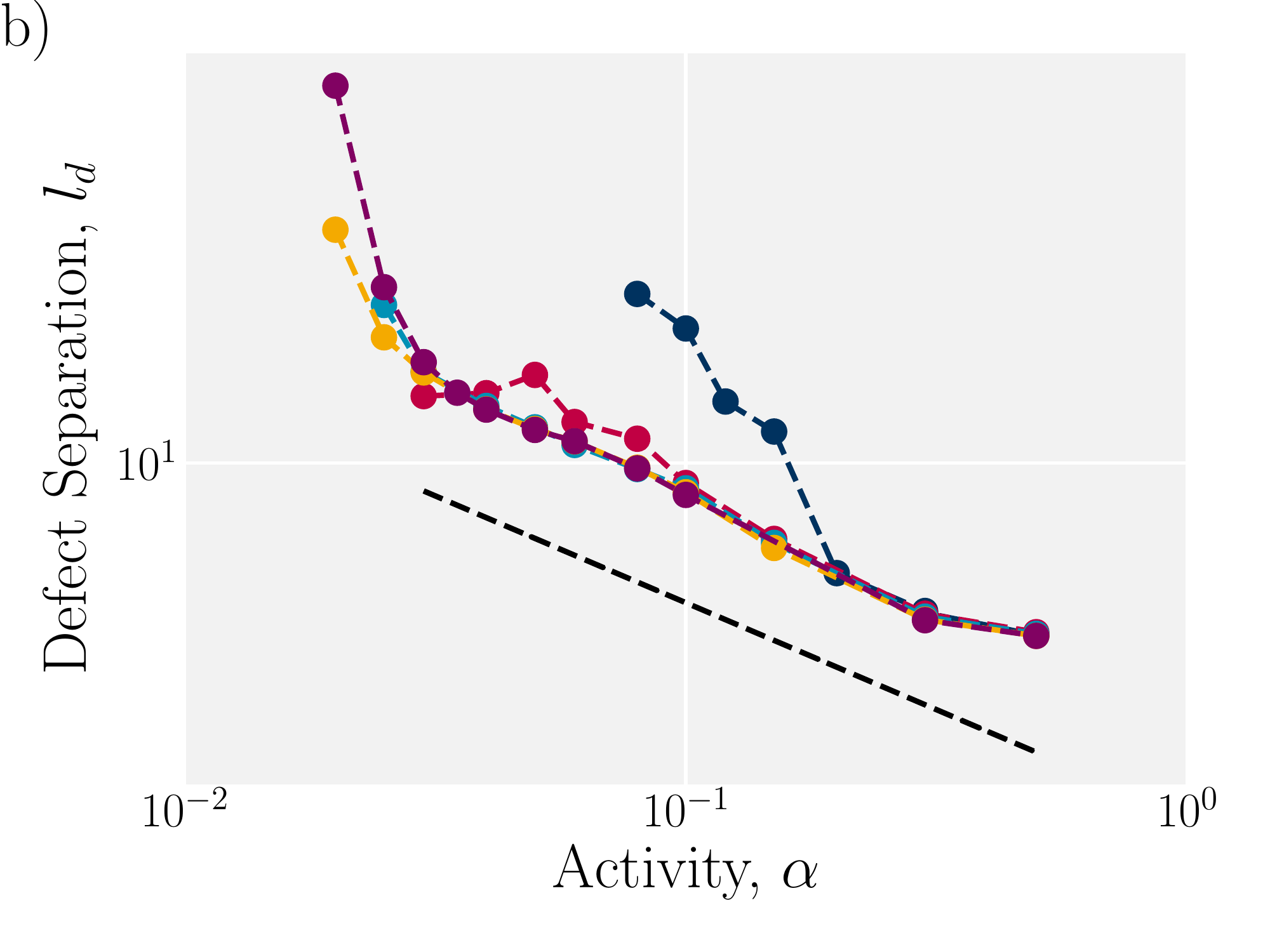}
        \label{fig:defSepAlign}
    \end{subfigure}
    \caption{\textbf{System size effects on defect separation.} 
    \textbf{(a)} Steady-state mean separation of defects $\lenDef = \rho_\text{d}^{-1/2}$ in the tumbling regime ($\lambda = 1/2$) as a function of activity for various system sizes $\lenSys$. 
    \textbf{(b)} Same as \fig{fig:nemOrderSysSize}a for shear-aligning active nematics ($\lambda = 2$). 
    The dashed lines indicate a scaling of $\lenDef\sim\alpha^{\mu}$ with $\mu=-1/2$.
    }
    \label{fig:nemOrderSysSize}
\end{figure}

\begin{figure}[tb]
     \centering
     \begin{subfigure}
         \centering
         \includegraphics[width=0.49\textwidth]{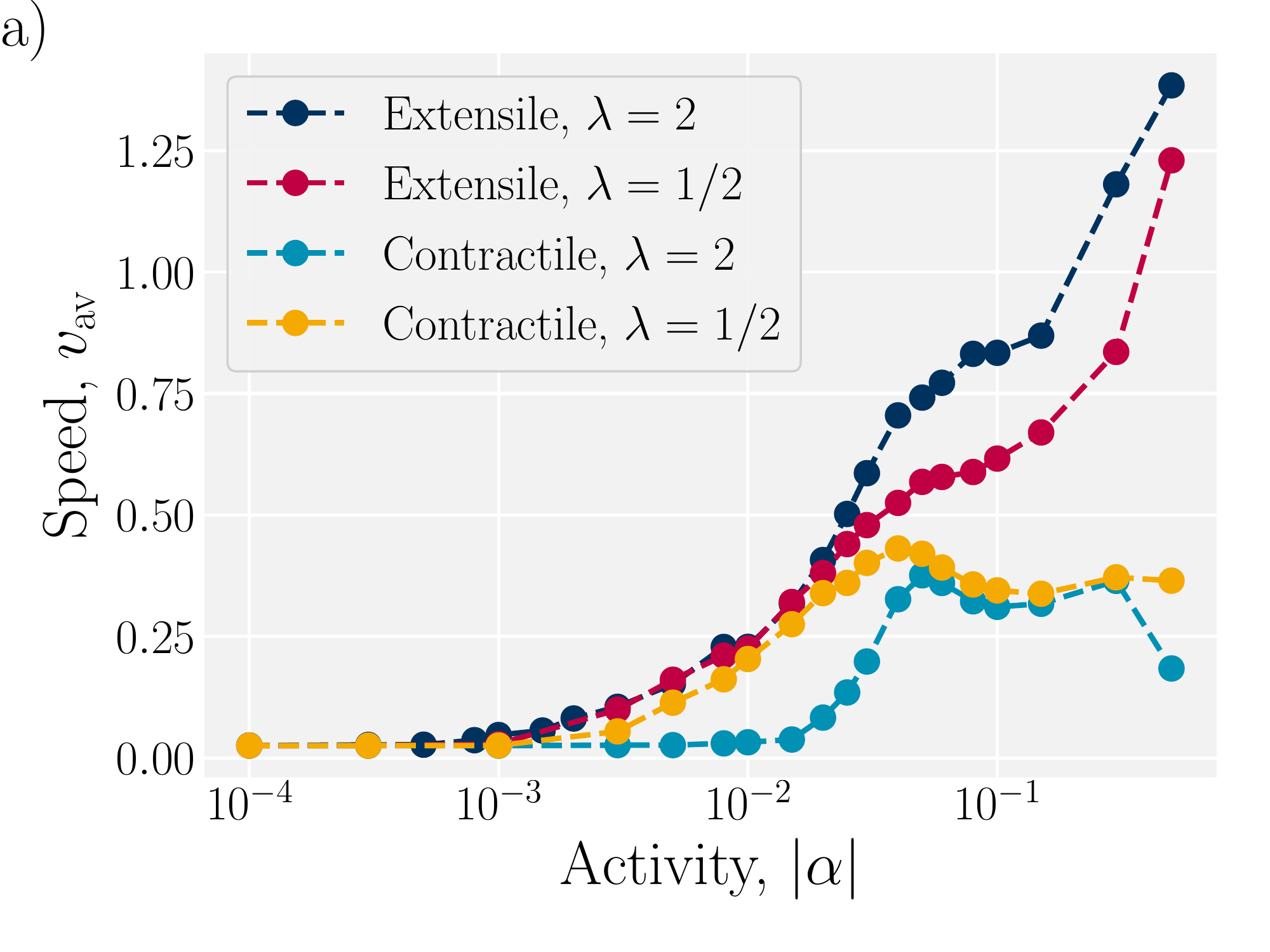}
         \label{fig:avSpeed}
     \end{subfigure}
     \begin{subfigure}
         \centering
         \includegraphics[width=0.49\textwidth]{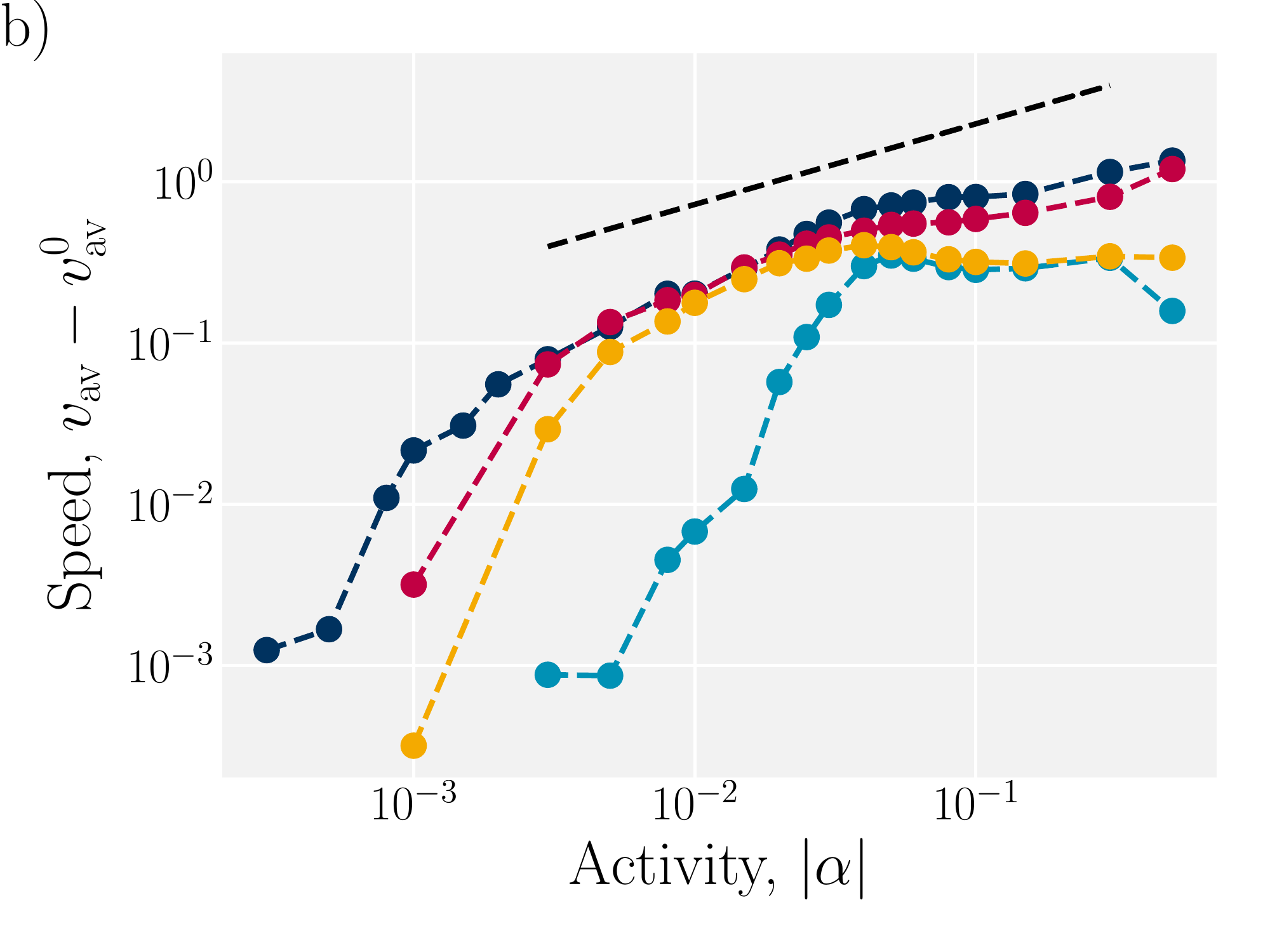}
         \label{fig:avSpeedLL}
     \end{subfigure}
     \caption{\textbf{Activity generates spontaneous flow.} 
     \textbf{(a)} Root mean square velocity for extensile and contractile active nematics in both the flow tumbling ($\lambda=1/2$) and shear-aligning ($\lambda=2$) regimes.  
     \textbf{(b)} Root mean square velocity on a log-log scale, with the ambient thermal speed subtracted. 
     The dashed line represents a scaling of $\velAv \sim \alpha^{\gamma}$ with $\gamma=1/2$. 
     }
     \label{fig:vel}
\end{figure}

One method to quantify coherent flow structure within the AN-MPCD fluid is to analyze the spatial autocorrelation functions
\begin{align}
    \corr{x}{R} &= \frac{ \av{\vec{x}\left(\vec{r}_0;t\right)\cdot\vec{x}\left(\vec{r}_0+R\hat{r};t\right)} }{ \av{\vec{x}\left(\vec{r}_0;t\right)\cdot\vec{x}\left(\vec{r}_0;t\right)} } ,
    \label{eq:corrDef}
\end{align} 
where $R$ is the separation distance between two points in the fluid and $\hat{r}$ is the radial unit vector. 
The average is over all points in space and time. 
The unspecified field $\vec{x}\left(\vec{r}_0;t\right)$ in \eq{eq:corrDef} might be the director or velocity fields, $\vec{x} \in \left\{ \vec{n}, \vec{v} \right\}$. 
Since this work focuses on the hydrodynamics that result from activity, the microscopic scale $R \leq R_{\kbt}$ within which thermal effects dominate correlation functions are removed from the analysis, and the remaining correlation function is renormalised such that $C_{xx}(R=0)=1$. 
Hydrodynamic correlations tend to initially decay and possess an anticorrelation well~\cite{Thampi2014RoySocA, Thampi2013PRL}. 
To measure of active length scales from autocorrelation functions, the decorrelation length is determined by fitting an exponential decay to the small-$R$ hydrodynamic region. 
The fit is preformed for the range $\corr{x} > e^{-1}\cdot \lim_{R\to\infty}\corr{x}$. 

Enstrophy spectra are employed when studying the spatial structure of traditional turbulence, and comparisons have been made to mesoscale active turbulence~\cite{Giomi2015PRX, Alert2020Nature}. 
For example, it has been found that Komogorov's universal $-5/3$ scaling exponent for inertial turbulence does not hold~\cite{Alert2020Nature}. 
The enstrophy $\Omega=\abs{\vec{\vor}\cdot\vec{\vor}}$ represents the magnitude of the vorticity $\vec{\vor}=\del\times\vec{\vel}$. 
In this work, the enstrophy spectra are computed via the radial Fourier transform of the vorticity correlation function
\begin{align}
	E_\Omega\left(k\right) &= \int_0^\infty \corr{\omega}{R} J_0(kR) R \,dR
	\label{eq:enstrophy}
\end{align}
where $J_0$ is the Bessel function of the first kind with order zero.

\subsection{Density Analysis}
A common feature of active particle-based models is their tendency to exhibit giant variations in local density~\cite{Chate2006PRL, PeshkovChate2014PRL, Bertin2009JPhysA, Bar2020AnnRev}. 
Nematic adaptations of the Vicsek model exhibit such giant number fluctuations~\cite{Chate2006PRL, GinelliChate2010PRL, Chate2019PRL}.
Furthermore, these have been observed in experimental and theoretical work on active-nematics~\cite{Ramaswamy2003EPL, Ramaswamy2007Science}.
As a particle-based algorithm for active nematics, AN-MPCD is expected to exhibit significant number fluctuations. 

In the typical case of equilibrium systems, density distributions are Gaussian due to the Central Limit Theorem~\cite{Ramaswamy2007Science}.
However, continuous energy injection can lead to non-Gaussian distributions.
Hence, a measure of a distribution's Gaussianity can be revealing.
A non-Gaussianity measure (NGM) can be defined to be
\begin{align} 
    \NGM &= \frac{d}{d+2} \frac{\Delta r^4}{\abs{\Delta r^2}} - 1
    \label{eq:NGM}
\end{align}
where $d=2$ is the dimension and $\Delta r^k$ is the $k$'th moment of the distribution\cite{Shendruk2020SciRep-Twitcher}. 
When a distribution is purely Gaussian $\NGM=0$. 
However, $\NGM > 0$ occurs when the the tails of the distribution stretch, and $\NGM < 0$ occurs when the tails contract relative to normal. 
Physically,  $\NGM \simeq 0$ holds for density distributions of systems in equilibrium, while active-particle models exhibit non-Gaussinity. 

To quantify the fluctuations in density, consider how the standard deviation of density, $\sigma_{\NCell}$, scales with the mean number density $\av{\NCell}$~\cite{Ramaswamy2007Science}. 
In equilibrium, one expects 
\begin{align}
    \label{eq:GNF}
    \sigma_{\NCell} \sim \av{\NCell}^\nu,
\end{align}
with the scaling $\nu=1/2$ given by the Central Limit Theorem. 
However, in systems of active particles, $\nu > 1/2$ is typical. 
As the scaling approaches $\nu \simeq 1$, an active system is said to exhibit \textit{giant number fluctuations}.
These fluctuations in density are predicted by theory~\cite{Ramaswamy2003EPL}, experiments~\cite{Ramaswamy2007Science} and observed in simulations~\cite{Shi2013NatComm}. 
The simulation domain is segmented into sub-domains in order to quantify these.
By averaging over these sub-domains and computing the mean and standard deviations, the scaling $\nu$ is obtained.

\begin{figure}[tb]
    \begin{subfigure}
        \centering
        \includegraphics[width=0.49\textwidth]{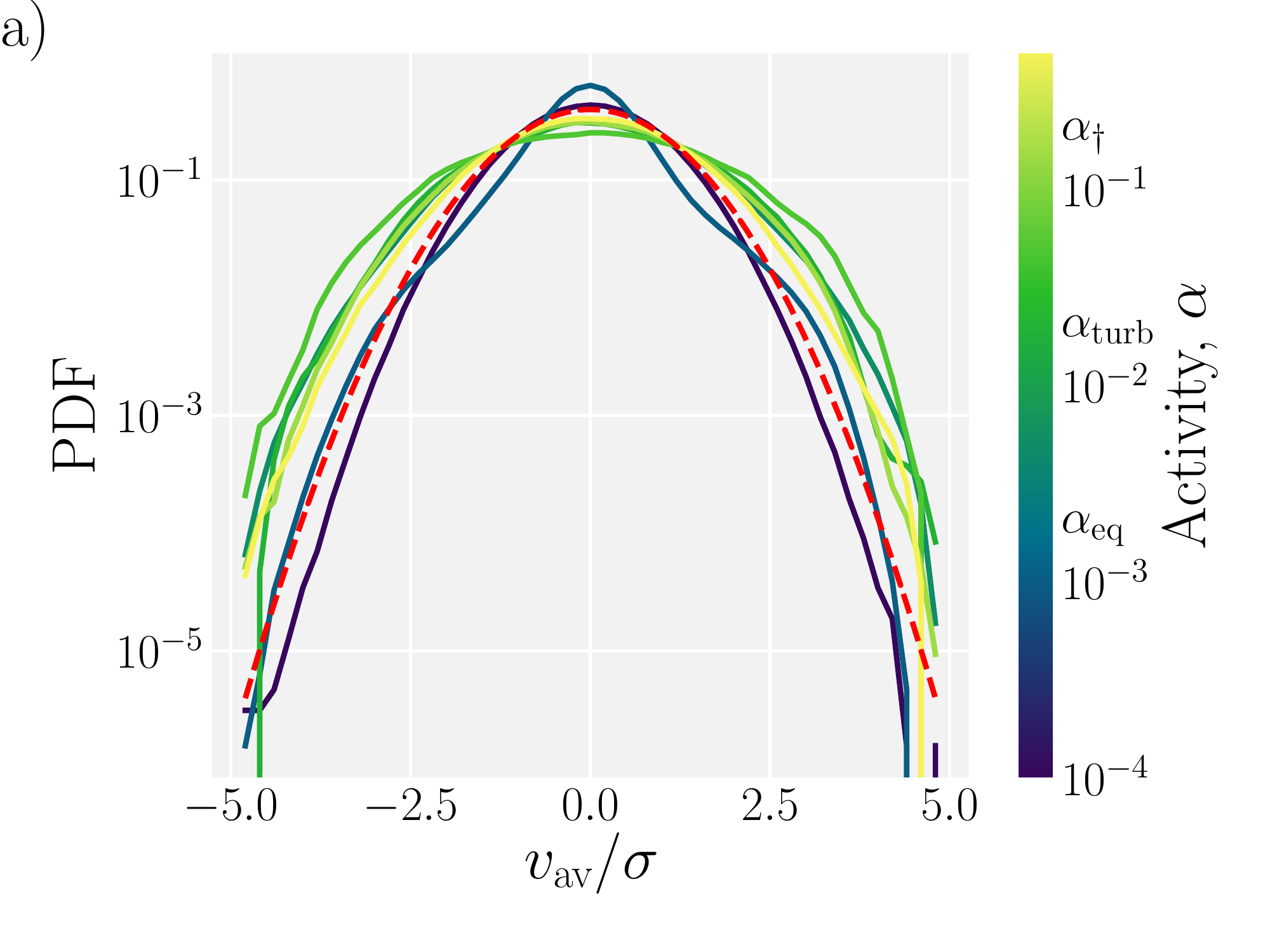}
        \label{fig:velDist}
    \end{subfigure}
    \begin{subfigure}
        \centering
        \includegraphics[width=0.49\textwidth]{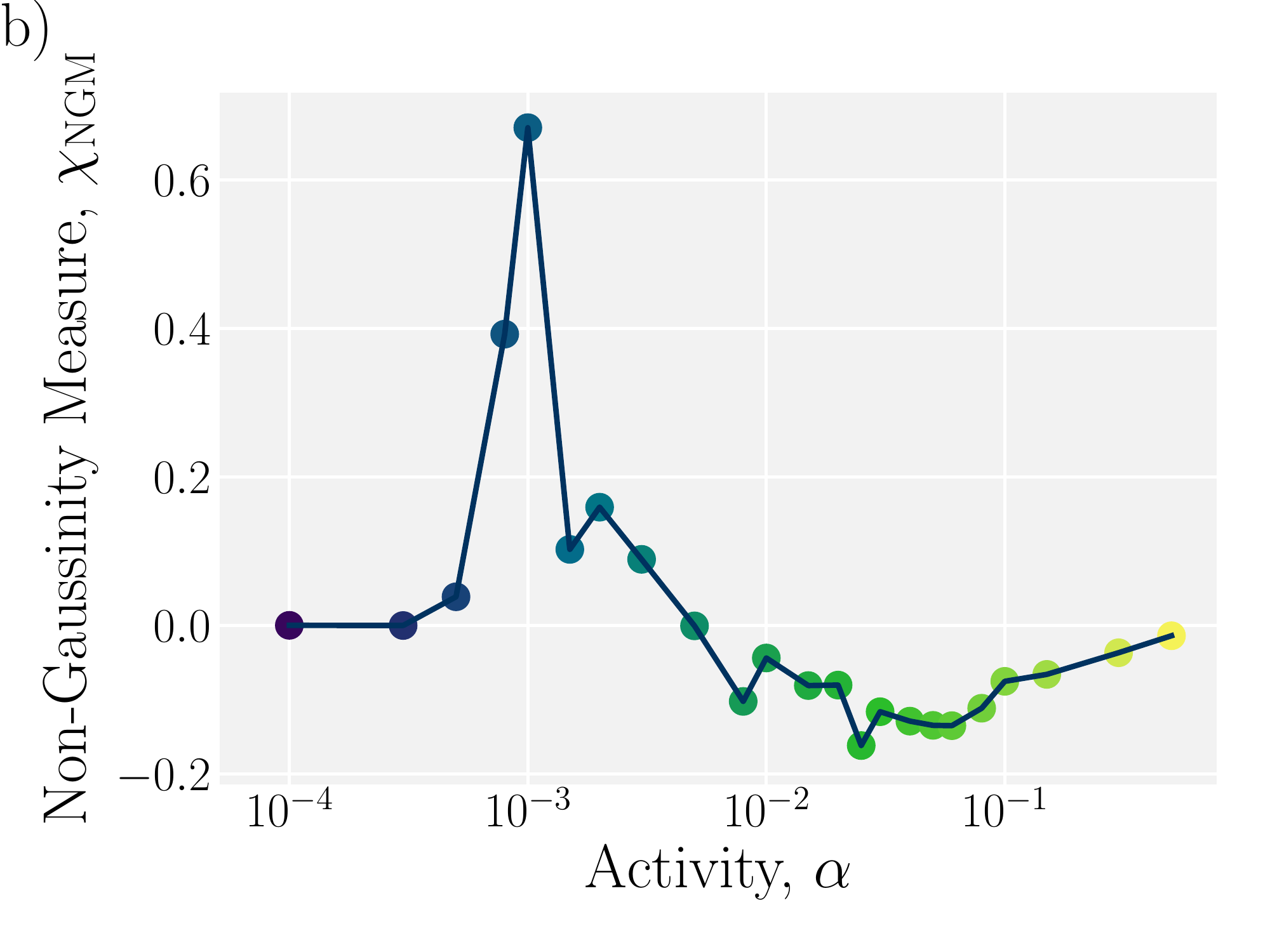}
        \label{fig:velDistNGM}
    \end{subfigure}
    \caption{\textbf{Activity broadens velocity distributions in AN-MPCD} 
    \textbf{(a)} Probability distributions of the local average of velocity components, rescaled by the standard deviation, for $\lambda=2$ and $\lenSys=200$. 
    The red dashed line is a standard Gaussian distribution.
    \textbf{(b)} Non-Gaussianity measure $\NGM$ of the distributions in \fig{fig:velProb}a. 
    }
    \label{fig:velProb}
\end{figure}

\section{Results}\label{sctn:results}
We now present the results of AN-MPCD simulations by first demonstrating that a hydrodynamic instability causes a proliferation of topological defects. 
The dependence of the emergent flows to the activity $\act$ are systematically explored by considering the structure of the flow fields. 
The results show that AN-MPCD possesses an operational active-turbulence-scaling regime, in which fully developed active turbulence exists with a characteristic emergent length scale $\lenAct$ that scales with activity.
The effects of activity on the local density are characterized and it is found that AN-MPCD exhibits giant-number fluctuations at sufficiently high activity. 

\subsection{Hydrodynamic Instability and Defect Unbinding}

Experiments~\cite{Dogic2015NatureMat, Dogic2019RSC, martinez2019} and continuum simulations~\cite{ThampiYeomans2014PRE, Yeomans2016NatComm} of active nematic systems exhibit activity-induced instabilities, which generate inhomogeneities in the nematic order. 
This leads to a steady-state population of topological defects and active turbulence~\cite{Thampi2014EPL}. 
The AN-MPCD algorithm reproduces this two-stage development of active turbulence through hydrodynamic instability and defect pair-creation. 

This processes is depicted by simulation snapshots in \fig{fig:pairCreation} from \movie{mov:defUnbindDir}.
The simulation is initialised in a fully ordered state with all nematogen orientations parallel to the $\hat{y}$-axis, \ie $\vec{u}_i=\pm\hat{y}$ $\forall \ i$ (\fig{fig:pairCreation}a) and the warmup period is eschewed. 
At early times the hydrodynamic instability manifests as kink walls --- narrow,  initially parallel, well-spaced, alternating regions of high bend deformation (\fig{fig:pairCreation}b). 
For extensile activity, fluctuations in bend produce active forces proportional to $\sim\alpha\del\cdot\tens{Q}$, which drive spontaneous flows along the kink walls. 
However, these self-driven streams exacerbate the inhomogeneities through the back flow, which in turn increases the active forces causing the instability~\cite{SimhaRamaswamy2002PhysicaA, VoituriezProst2005EPL, Ramaswamy2007NJP}. 

For sufficiently high activity, this hydrodynamic-bend instability is responsible for the formation of topological defect pairs (\fig{fig:pairCreation}c).
Beyond a certain point, the total distortion free energy cost along a kink wall is higher than the cost of two oppositely charged defects and a pair is spontaneously formed~\cite{Thampi2014EPL}. 
Although there is an elastic restoring force, the $+1/2$ defect is a self-propelled quasi-particle and it moves along the kink wall, becoming unbound from the non-motile $-1/2$ defect~\cite{Thampi2014EPL, Marchetti2018PRL} (\fig{fig:pairCreation}d). 

A pair-creation event is illustrated in detail in the top row of \fig{fig:Creaton-Annihilation}. 
Along a kink wall, a pair of $\pm1/2$ defects are formed, conserving the net topological charge. 
The motility of $+1/2$ defects causes the pair to unbind. 
As the motile $+1/2$ defect travels along the kink wall, it alleviates the deformation free energy, leaving the region of uniform order in its trail (\fig{fig:Creaton-Annihilation}; top row) but generating local vorticity, which in turn perturbs nearby kink walls, disordering the orientation of future defect pair unbinding events giving rise to active turbulence (see \fig{fig:pairCreation}c-d).
Though activity generates defects, they come with an inherent free energy cost due to the deformation they induce in the director field, and as such they annihilate to minimise the deformation cost (\fig{fig:Creaton-Annihilation}; bottom row).
As a $+1/2$ defect approaches a $-1/2$ defect, they annihilate and ordering the local region. 
Continual creation and annihilation leads to a steady-state defect population. 

\begin{figure}[tb]
     \centering
     \begin{subfigure}
         \centering
         \includegraphics[width=0.4\textwidth]{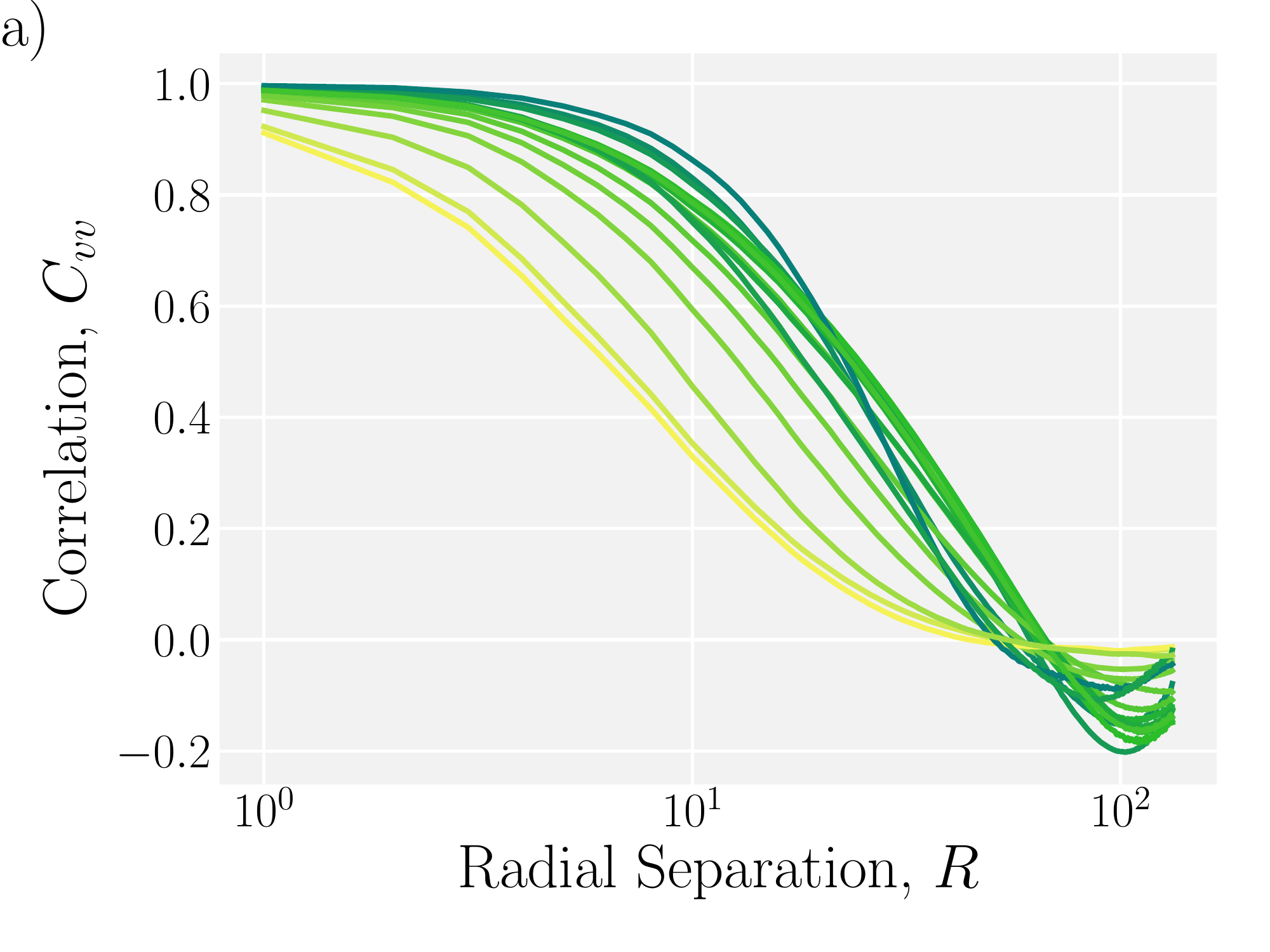}
         \label{fig:velCorr}
     \end{subfigure}
     \begin{subfigure}
         \centering
         \includegraphics[width=0.4\textwidth]{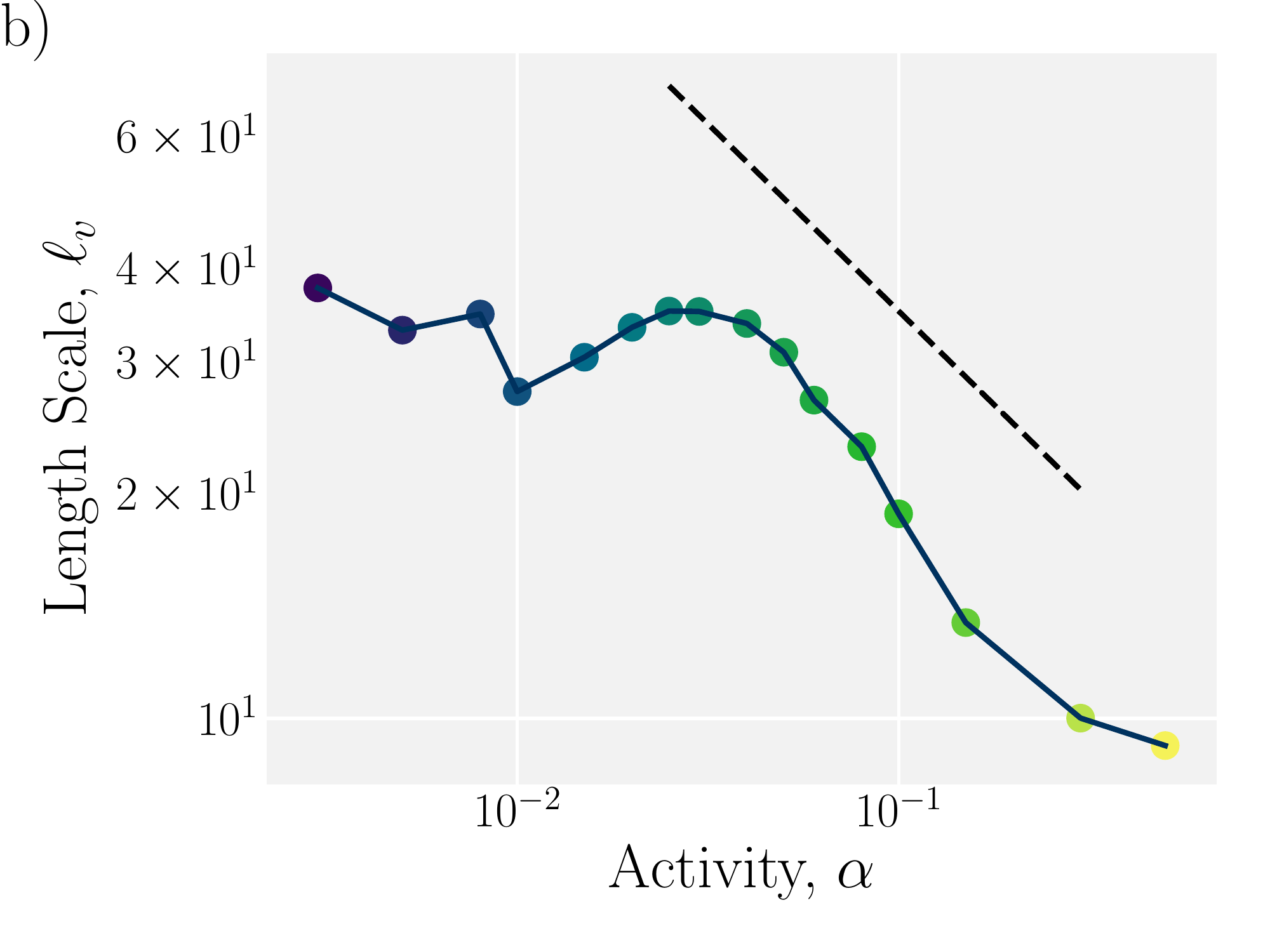}
         \label{fig:velLength}
     \end{subfigure}
     \begin{subfigure}
         \centering
         \includegraphics[width=0.4\textwidth]{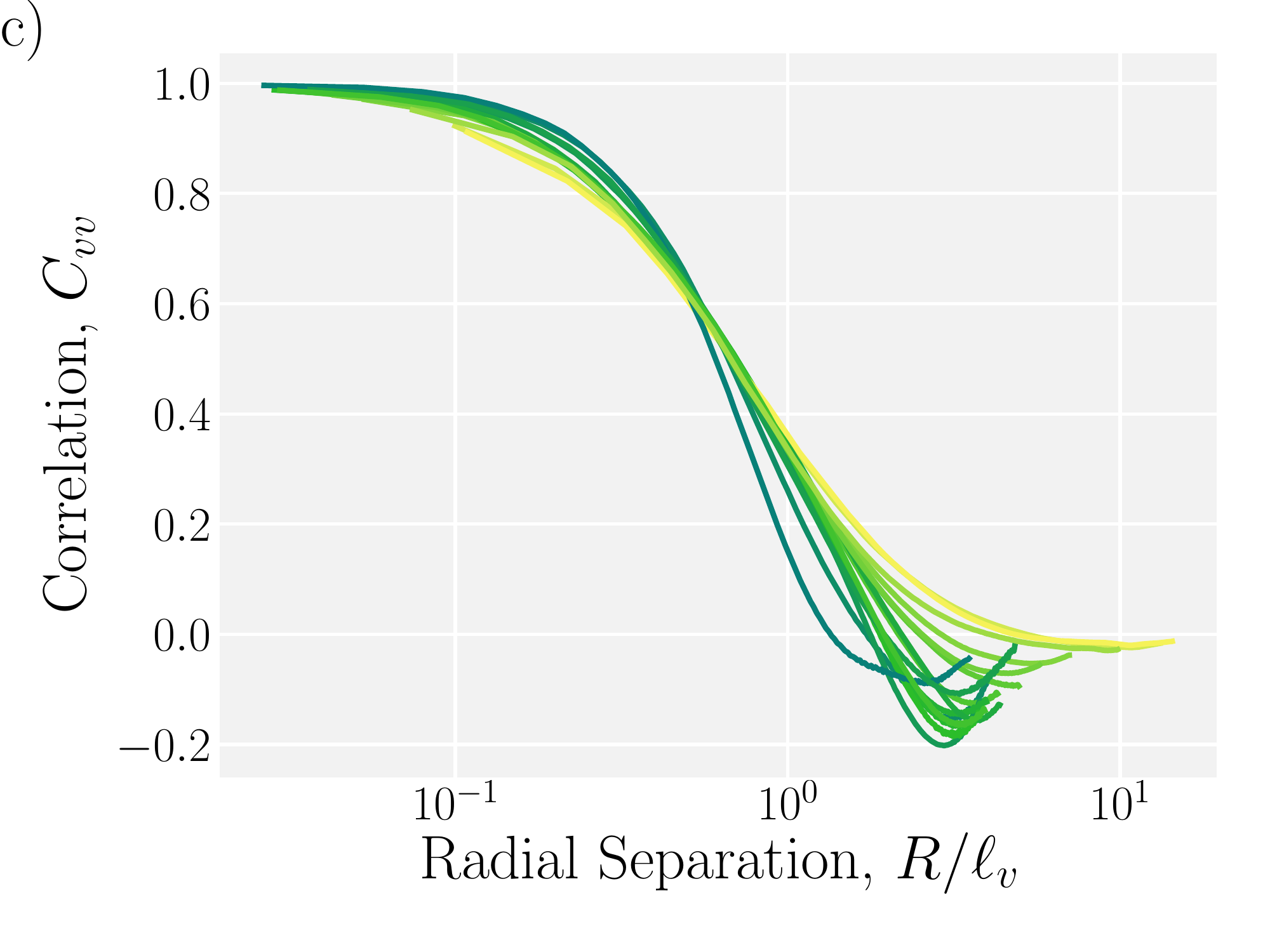}
         \label{fig:velCorrScaled}
     \end{subfigure}
     \caption{\textbf{Spatial structure of active turbulence.} 
     \textbf{(a)} Radial hydrodynamic autocorrelation functions of velocity for various activities above $\actMIN$. Colors match previous figures. 
     \textbf{(b)} The velocity length scales as computed from the correlation functions in \fig{fig:turb}a. 
     The dashed line indicates a scaling of $\lenVel\sim\alpha^{\mu}$ with $\mu-1/2$.
     \textbf{(c)} Scaling the correlation functions in  \fig{fig:turb}a by the corresponding length scale in  \fig{fig:turb}b.
     }
     \label{fig:turb}
\end{figure}

\subsection{Steady-State Defect Population}
To assess the degree of nematic ordering within the steady-state AN-MPCD fluid, the spatial autocorrelation of the director field $\corr{n}{R}$ is considered. 
For the lowest activities considered ($\act\lesssim \actMIN=10^{-3}$), the correlation remains high even as $R\to\lenSys$ (\fig{fig:nemOrder}a). 
In these cases, activities less than $\actMIN$ are negligible and the injected energy can be absorbed by the thermostat so that the system remains globally ordered.
This is seen in \movie{mov:a0.0008dir} where the nematic director field remain oriented primarily in the $\hat{y}$-axis, with slight thermal fluctuations. 
For the low activity regime ($\actMIN \lesssim \act \lesssim \actMin$ where $\actMin\simeq10^{-2}$), the director field decorrelates somewhat but does maintain long-range correlations/anti-correlations, which represent the persistent existence of kink walls with a characteristic separation length (\movie{mov:a0.0080dir}). 
Only once $\act \gtrsim \actMin$ do the correlation functions fully decorrelate to $\lim_{R\to\infty}\corr{n}{R} \to 2/3$ and dip below the long-range value, as observed in continuum simulations of fully formed active turbulence (\movie{mov:a0.0800dir})~\cite{Thampi2014RoySocA}. 
The activity $\actMin=10^{-2}$ represents the lower threshold between fully developed active turbulence and kink walls for this system size. 
In the active turbulence regime, $\actMin \lesssim \act$, we find that the correlations belong to the same class of functions by rescaling the separation by the nematic decorrelation length (\fig{fig:nemOrder}b). 
Rescaling collapses the curves for $\act \gtrsim \actMin$.

Consistent with the measured correlation functions, the number of defects is found to be non-zero only for sufficiently strong activity $\act \gtrsim \actMin$, where the minimum activity to observe turbulence is $\actMin = 2.5\times10^{-2}$ for $\lenSys=200$ (\fig{fig:defSep}). 
Above $\actMin$ the mean defect density $\rho_\mathrm{d}$ rises rapidly and the root mean squared separation between $\pm1/2$ defects, $\lenDef=\rho_\text{d}^{-1/2}$, decreases accordingly (\fig{fig:defSep}, inset).
The scaling $\lenDef \sim \act^{\mu}$ with $\mu=-1/2$ is expected for a range above $\actMin$ (\eq{eq:actLength}) and the simulation results are found to match the expected scaling within the fully formed turbulence region $\actMin \leq \act \leq \actMax$ (\fig{fig:defSep} inset). 
Thus, continual creation and annihilation leads to a steady-state defect population in AN-MPCD characterized by the competition between activity and nematic elasticity (\eq{eq:actLength}). 

To assess system size effects, larger and smaller systems $\lenSys\in\{400,200,100,50,25\}$ for both tumbling ($\lambda = 1/2$ in \fig{fig:nemOrderSysSize}a) and shear-aligning ($\lambda = 2$ in \fig{fig:nemOrderSysSize}b) behaviours are simulated. 
In both cases, the system size shifts the activity at which defect pairs unbind and the range of activities for which the active-turbulence scaling law holds is extended for larger system sizes. 

\subsection{Spontaneous Flow}

The previous section demonstrates that the AN-MPCD algorithm reproduces the hydrodynamic instability, topological defect propagation, the self-motility of $+1/2$ defects and the expected scaling of the characteristic length scale as measure by the mean separation between defects. 
This section explores the spontaneous flows and confirms that AN-MPCD reproduces the properties of fully developed active turbulence across a range of activities above $\actMin$. 

The magnitude of the active flows is quantified by measuring the steady-state, spatiotemporally averaged, root mean squared velocity $\velAv$ (\fig{fig:vel}a). 
Since AN-MPCD is a thermalized method, $\lim_{\act\to0}\velAv\neq0$ and a small-but-non-zero value of $\velAv$ is measured below $\actMIN=10^{-3}$ (\movie{mov:a0.0008vel} and \fig{fig:vel}a). 
Above $\actMIN$ the average speed rises rapidly --- more rapidly than predicted by \eq{eq:actVelocity} (\fig{fig:vel}b). 
This is consistent with \fig{fig:nemOrder}: 
Between $\actMIN$ and $\actMin$, the activity is sufficient to generate kink walls (\movie{mov:a0.0080dir}) and spontaneous flow (\movie{mov:a0.0080vel}) but insufficient to produce defects and active turbulence. 
Thus, the scale of the characteristic speed rises faster than when active turbulence is fully formed. 
Active turbulence is fully formed at $\actMin\simeq10^{-2}$ (\movie{mov:a0.0800vel}) and the scaling $\velAv \sim \act^\gamma$ is seen to be in agreement with the expected value of $\gamma=1/2$ (\eq{eq:actVelocity}), shown as a dashed line in \fig{fig:vel}b, for the region $\actMin \lesssim \act$. 

As activity increases, $\velAv$ does too, representing the broadening of the distribution of velocities. 
However, this is not the only effect; the probability density function also changes shape (\fig{fig:velProb}a). 
For the lowest activities ($\act\lesssim\actMIN$), the distributions of the $\hat{x}$- and $\hat{y}$-components are Gaussian. 
For larger activities ($\act\gtrsim\actMIN$), the distribution exhibits longer tails and the kurtosis is larger than for normal distributions, which is indicative of the out of equilibrium behaviour and correlated flow fields. 
To quantify the degree of non-equilibrium behaviour, a non-Gaussianity measure ($\NGM$) for the velocity distributions is considered in \fig{fig:velProb}b.
For the lowest activities, $\NGM=0$, corresponding to a true Gaussian distribution for velocity; however, this quickly peaks around $\act\simeq\actMIN$.
It is worth noting that our non-Gaussinity measure is directly related to the Binder cumulant~\cite{binder2021}, and the peak is indicative of discontinuous onset of activity-induced flows. 
The positive peak in $\NGM$ corresponds to the single narrower distribution in \fig{fig:velProb}a.  
As activity increases past $\actMIN$ the non-Gaussinity measure drops to negative values, and remains negative throughout the turbulence scaling region $\actMin \lesssim \act \lesssim \actMax$, corresponding to the broadening velocity distributions in \fig{fig:velProb}a.
Finally, for $\actMax \lesssim \act$, the non-Gaussinity measure increases towards $\NGM\to0$, indicating the breakdown of the algorithm. 

To quantify the flow structures responsible for the non-Gaussianity, the velocity autocorrelation functions $\corr{v}{R}$ are measured (\fig{fig:turb}a). 
By fitting an exponentially decaying function, the characteristic velocity lengths are obtained (\fig{fig:turb}b). 
For $\act\lesssim\actMIN$, activity is insufficient to induce hydrodynamic effects.
Just as in \fig{fig:vel}, for $\actMIN\lesssim\act\lesssim\actMin$ hydrodynamic effects are apparent, in that there is a characteristic hydrodynamic length scale.
However, there is still insufficient activity to drive active turbulence at these low activities and so the decorrelation length does not yet scale with activity. 
Only once $\actMin \gtrsim \act$ does the velocity decorrelation length scale decrease with activity. 
In this regime, the decrease in length scale is seen to correspond to the ideal expectations $\lenVel \sim \alpha^ \mu$ where $\mu = -1/2$ (\eq{eq:actLength}), shown as a dashed line. 
Comparing to \fig{fig:defSep}, the director and velocity decorrelation length scale both scale as expected within the fullly formed active turbulence regime, $\act \gtrsim \actMin$. 
Correspondingly, the measured length scale collapses the velocity-velocity correlation lengths in the fully formed active turbulence regime of $\act \gtrsim \actMin$ (\fig{fig:turb}c). 


\begin{figure}[tb]
     \centering
     \begin{subfigure}
         \centering
         \includegraphics[width=0.49\textwidth]{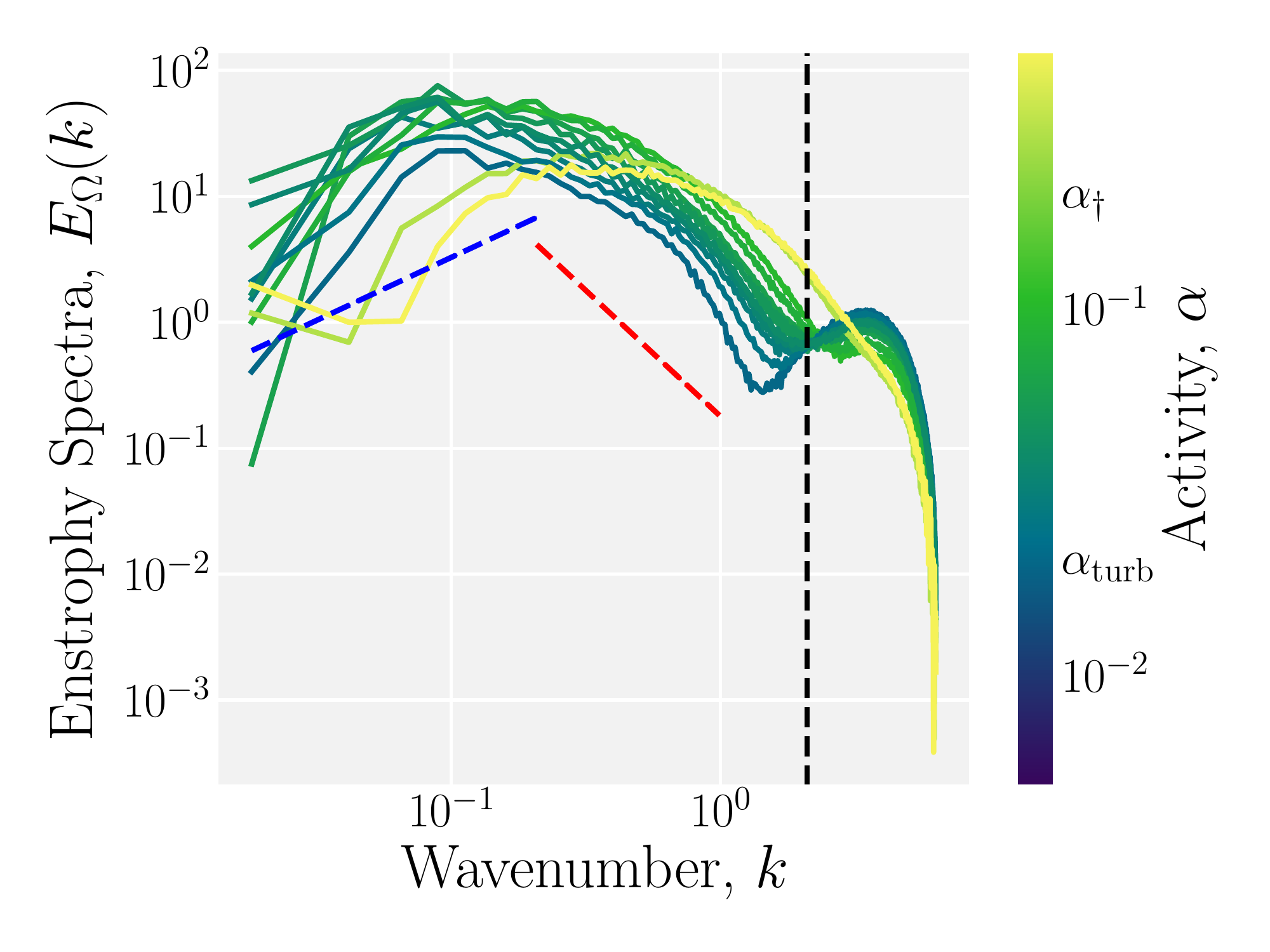}
         \label{fig:vorSpectra}
     \end{subfigure}
     \caption{\textbf{Enstrophy spectral structure of active turbulence.} 
     Enstrophy spectra for $\lenSys=400$. 
     Scalings of $\sim k^1$ at low wave numbers (blue dashed line) and $\sim k^{-2}$ at intermediate wave numbers (red dashed line) guide the eye. 
     At the highest wave numbers there exists a secondary structure due to the MPCD cell size $a$, which appears because vorticity is computed on the MPCD lattice. 
     The wave number at which vorticity computed is marked as the vertical black dashed line.
     }
     \label{fig:spect}
\end{figure}

\begin{figure*}[tb]
     \centering
     \begin{subfigure}
         \centering
         \includegraphics[width=0.49\textwidth]{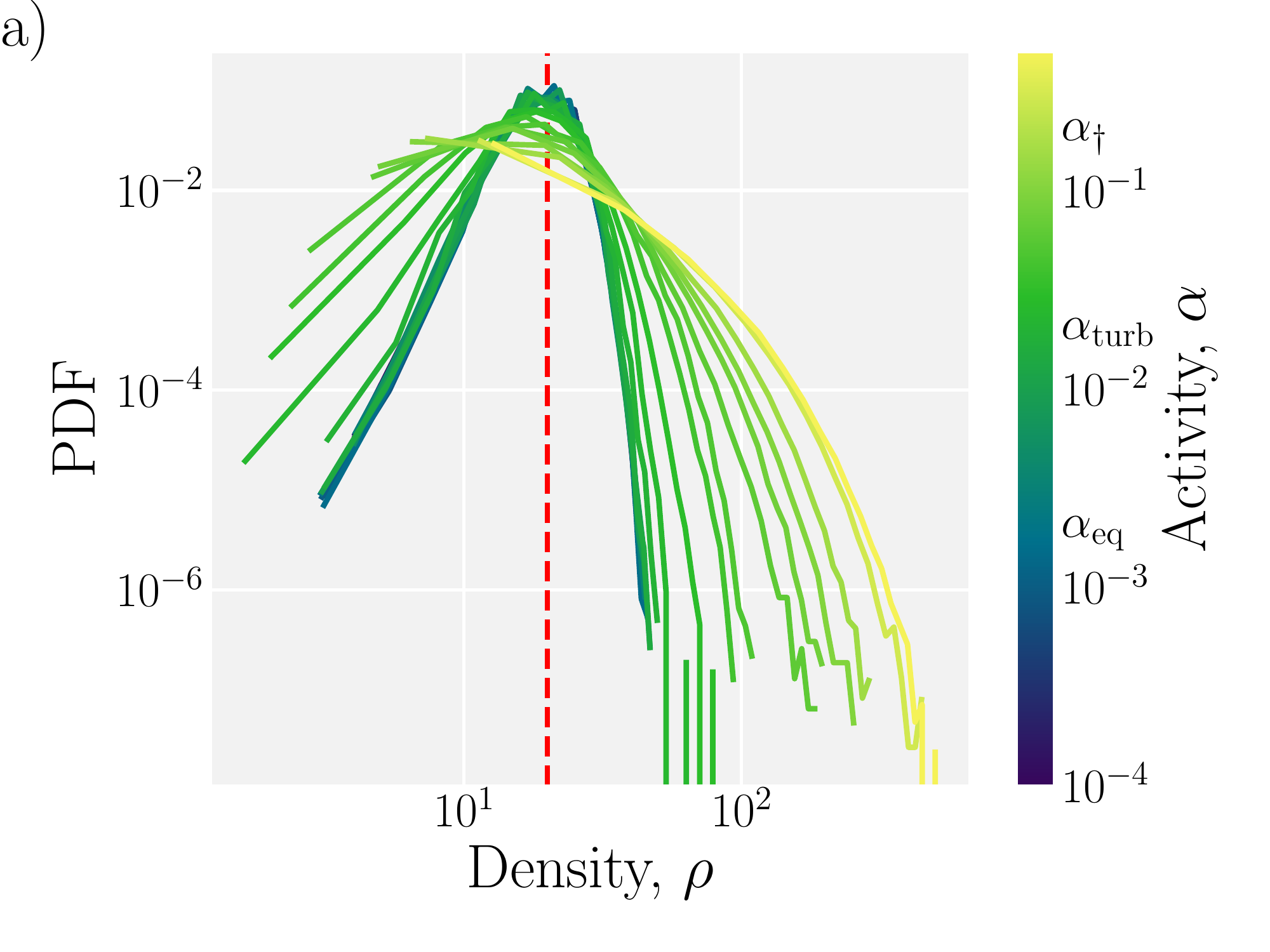}
         \label{fig:distDens}
     \end{subfigure}
     \begin{subfigure}
         \centering
         \includegraphics[width=0.49\textwidth]{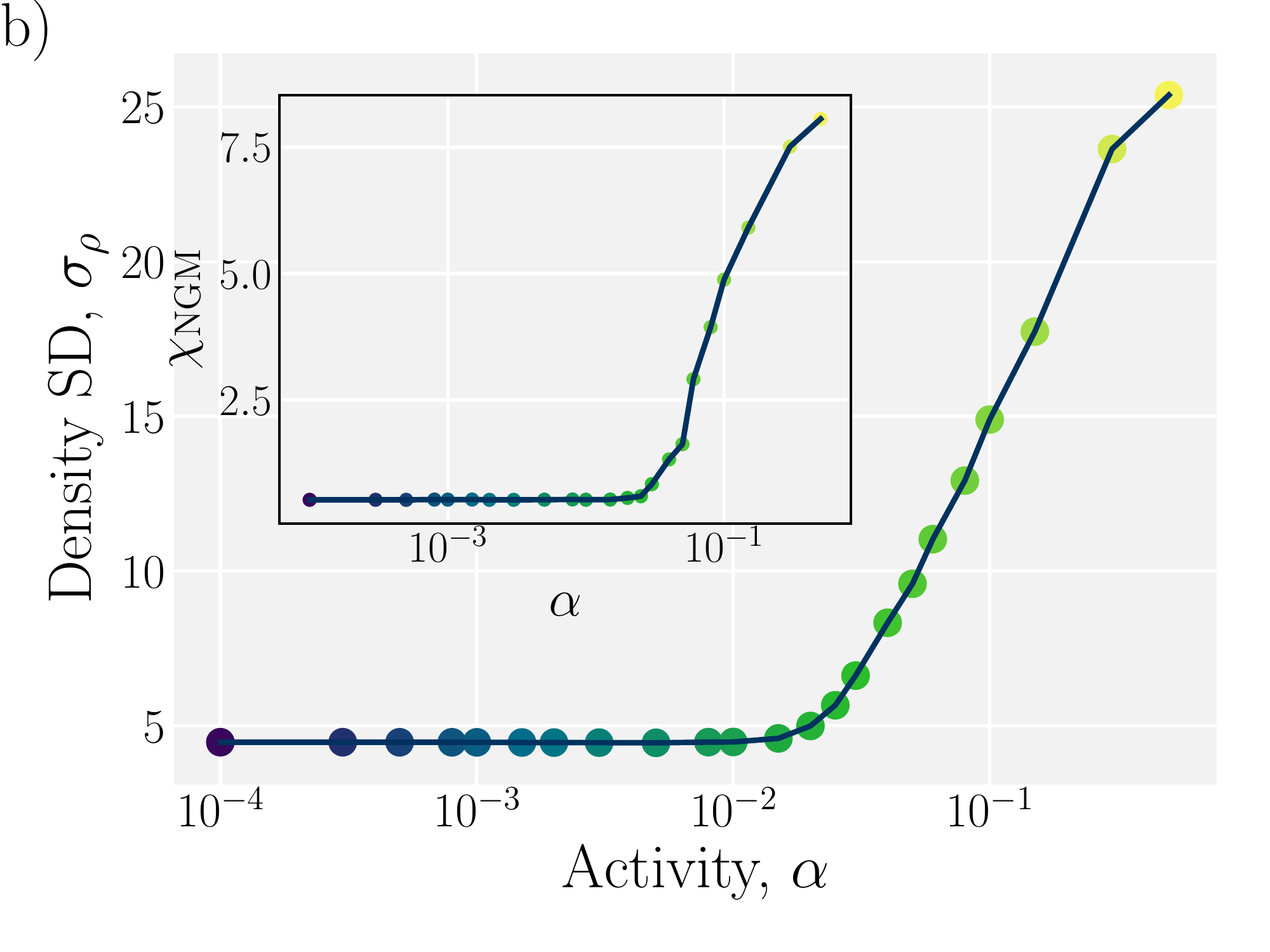}
         \label{fig:densStDev}
     \end{subfigure}
     \begin{subfigure}
         \centering
         \includegraphics[width=0.49\textwidth]{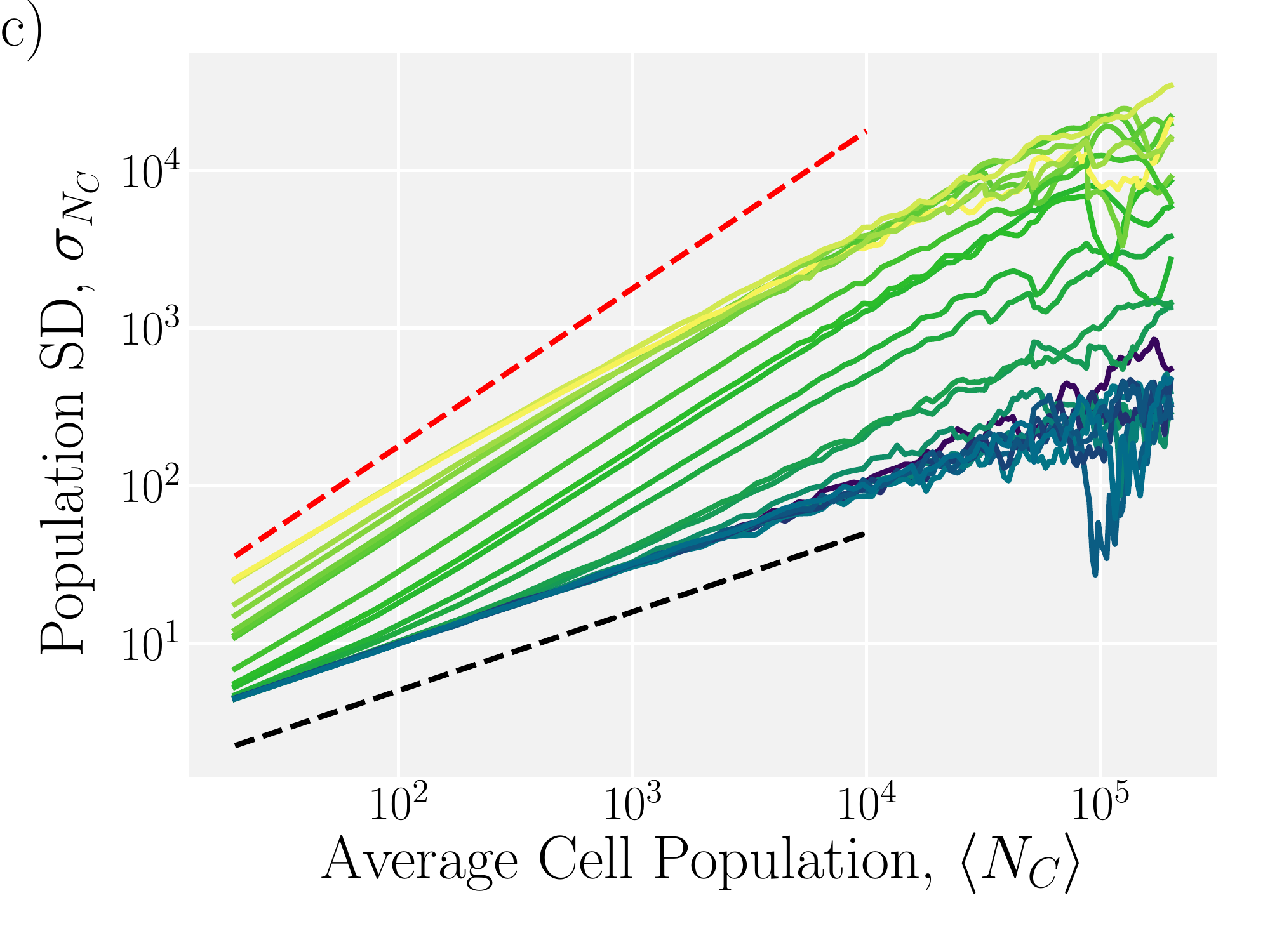}
         \label{fig:GNF}
     \end{subfigure}
     \begin{subfigure}
         \centering
         \includegraphics[width=0.49\textwidth]{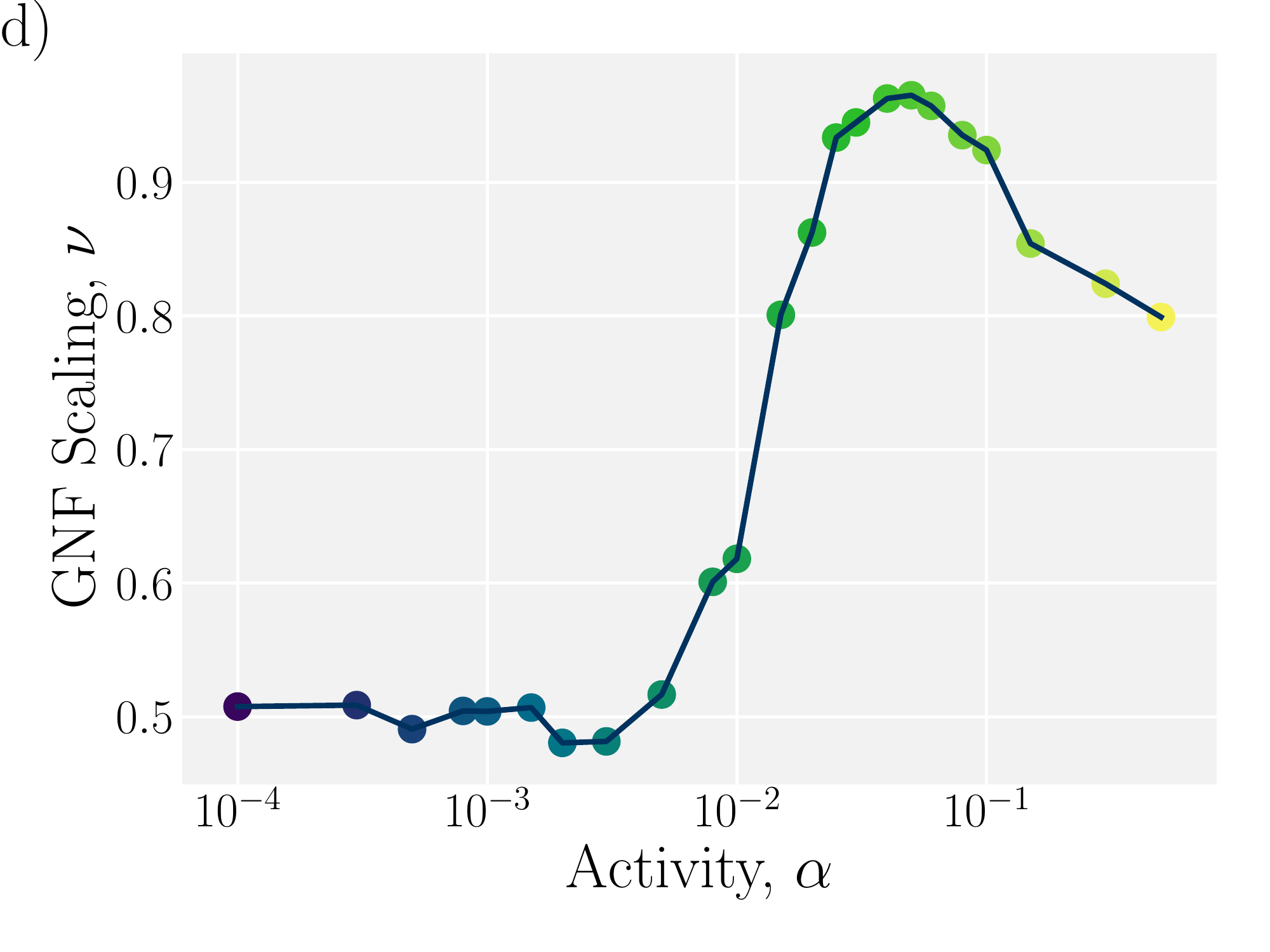}
         \label{fig:GNF_scaling}
     \end{subfigure}
     \caption{\textbf{AN-MPCD exhibits giant-number fluctuations.} 
     \textbf{(a)} Probability density functions (PDF) of the cell density. The average cell density $\av{\NCell}=20$ is shown as a dashed red line. 
     \textbf{(b)} Standard deviation of the cell density as a function of activity. 
     \textbf{(inset)} Non-Gaussinity measure $\NGM$ of cell density PDF \fig{fig:dens}a. 
     \textbf{(c)} Standard deviation of population shown as a function of the average cell population. 
     Scalings of $\sigma_{\NCell} \sim \av{\NCell}^\nu$ with $\nu=1/2$ (black dashed line) and $\nu=1$ (red line). 
     \textbf{(d)} The scaling exponent $\nu$ of \fig{fig:dens}c measured at the low cell population limit. 
     }
     \label{fig:dens}
\end{figure*}

Active nematics do not possess the scale invariance that inertial turbulence does~\cite{Yeomans2012PNAS, Frey2015PNAS, UrzayYeomans2017JFM, Alert2020Nature}. 
This is because energy is dissipated at the scale it is injected in, impeding energy cascades. 
In particular, the length scale from \eq{eq:actLength} represents the characteristic vortex size and so considering the amount of enstrophy as a function of wave number (\eq{eq:enstrophy}), reveals the spatial structure of active turbulence. 
At small wave numbers, the enstrophy rises with wave number rise $\sim k$ (\fig{fig:spect}); while at intermediate wave numbers, the enstrophy decreases as $\sim k^{-2}$ (\fig{fig:spect}). 
This change of sign in the scaling of the enstrophy spectra is expected to occur at $k > k_\act$, the wave number representing the characteristic vortex size~\cite{Giomi2015PRX, Alert2020NaturePhys}. 
The vortex structure results in a non-monotonic enstrophy spectra with a maximum corresponding to the characteristic vortex diameter. 
At the largest wave numbers, the signs of the MPCD cell discretization size $a$ are apparent as a secondary peak for $k\geq  k_a\approx2\pi/3a$ (\fig{fig:spect}). 
This is because nearest-neighbour MPCD cells are used to calculate the local vorticity. 

\begin{figure}[tb]
    \centering
    \begin{subfigure}
        \centering
        \includegraphics[width=.95\linewidth]{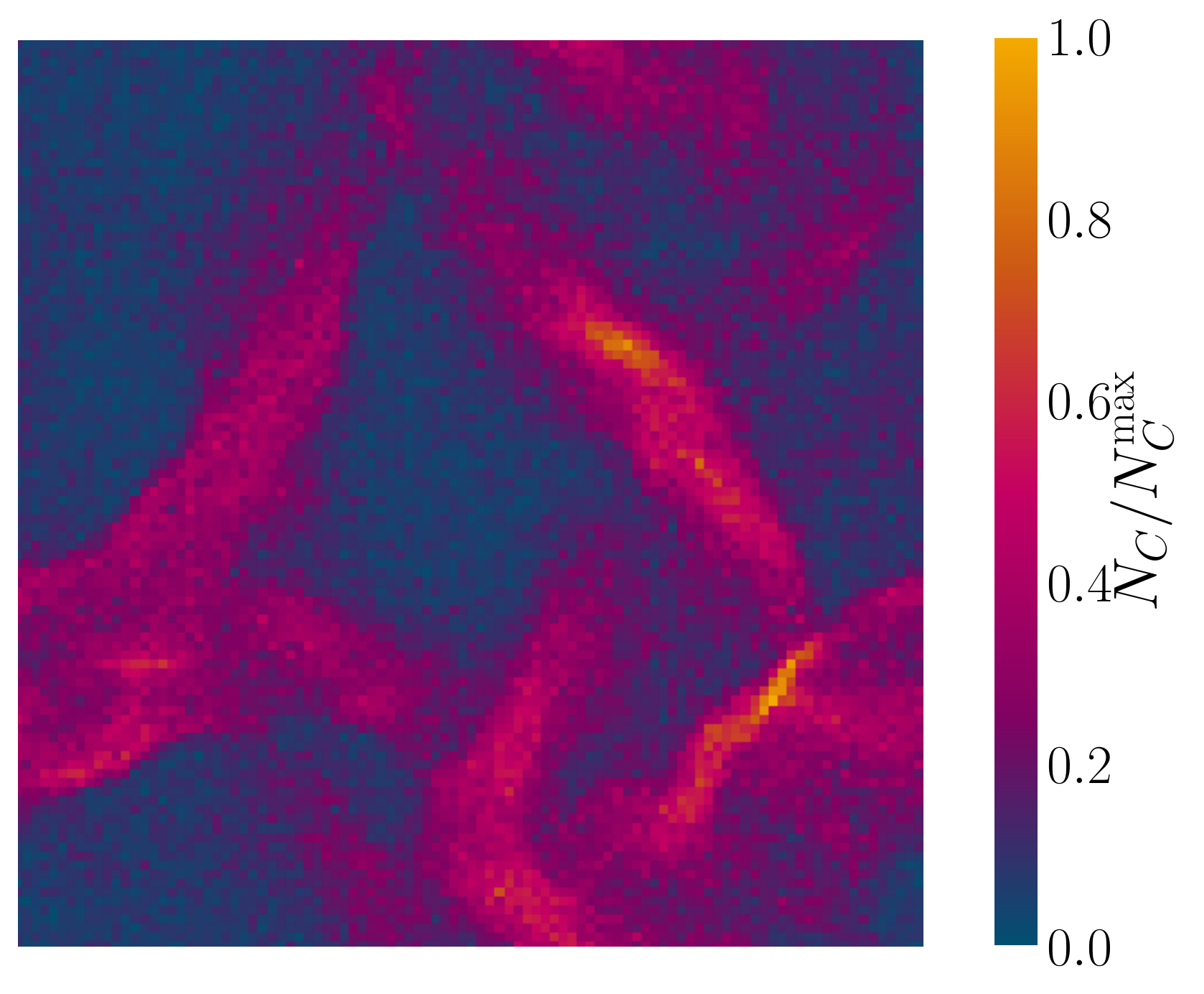}
    \end{subfigure}
    \caption{
        \textbf{Density Fluctuations in Active-Nematic MPCD.}
        Snapshot of the cell density field of an AN-MPCD system ($\lenSys=100$), $\act=0.08$, and average cell density $\av{\NCell}=20$ from \movie{mov:a0.0800dens}. 
        $N_C^\mathrm{max}$ is the instantaneous maximum cell population.
        AN-MPCD exhibits high density bands within a dilute nematic gas.
    }
     \label{fig:densityFluctuations}
\end{figure}

These results demonstrate the AN-MPCD algorithm generates spontaneous flows for sufficiently large activities. 
The scale of these flows increases rapidly at first, when the ratio of system size and activity are insufficient for fully developed turbulence, but for sufficient activity the characteristic velocity grows according to the theoretically expected scaling. 
The autocorrelation functions and enstrophy spectra demonstrate that AN-MPCD reproduces the essential properties of fully developed active turbulence with flow structures that statistically match expectations. 
However, AN-MPCD is a point particle-based mesoscale method and, as such, also possesses properties expected of active particle models. 

\subsection{Density Fluctuations}

As a mesoscale algorithm between the continuum and microscopic limits, AN-MPCD exhibits both the active turbulence predicted by continuum approaches and the density fluctuations of particle models. 
Since MPCD is composed of point-particles and obeys an ideal gas equation of state~\cite{Kapral2008AdvChemPhys-MPCD, GompperIhle2009Book-MPCD, Zantop2021JCP, Eisenstecken2018EPL}, it is a compressible fluid, as is illustrated by the probability distribution of the density (\fig{fig:dens}a). 
In the low activity limit $\act \ll \actMIN$, the fluid possesses a relatively broad distribution of particles with a mean $\rho=\av{\NCell}/a^2=20$ and a standard deviation $\sigma_{\NCell} \simeq 4.5$ (\fig{fig:dens}a, b). 
At the lowest activities, the mean coincides with the mode at $\NCell=20$, and the distribution is relatively Gaussian, with a slight positive skew. 
At these low activities, the distribution does not change as a function of activity; however, once $\act\geq\actMin$, the standard deviation begins to rise (\fig{fig:dens}b). 

\begin{figure}[tb]
    \centering
    \begin{subfigure}
        \centering
        \includegraphics[width=0.49\textwidth]{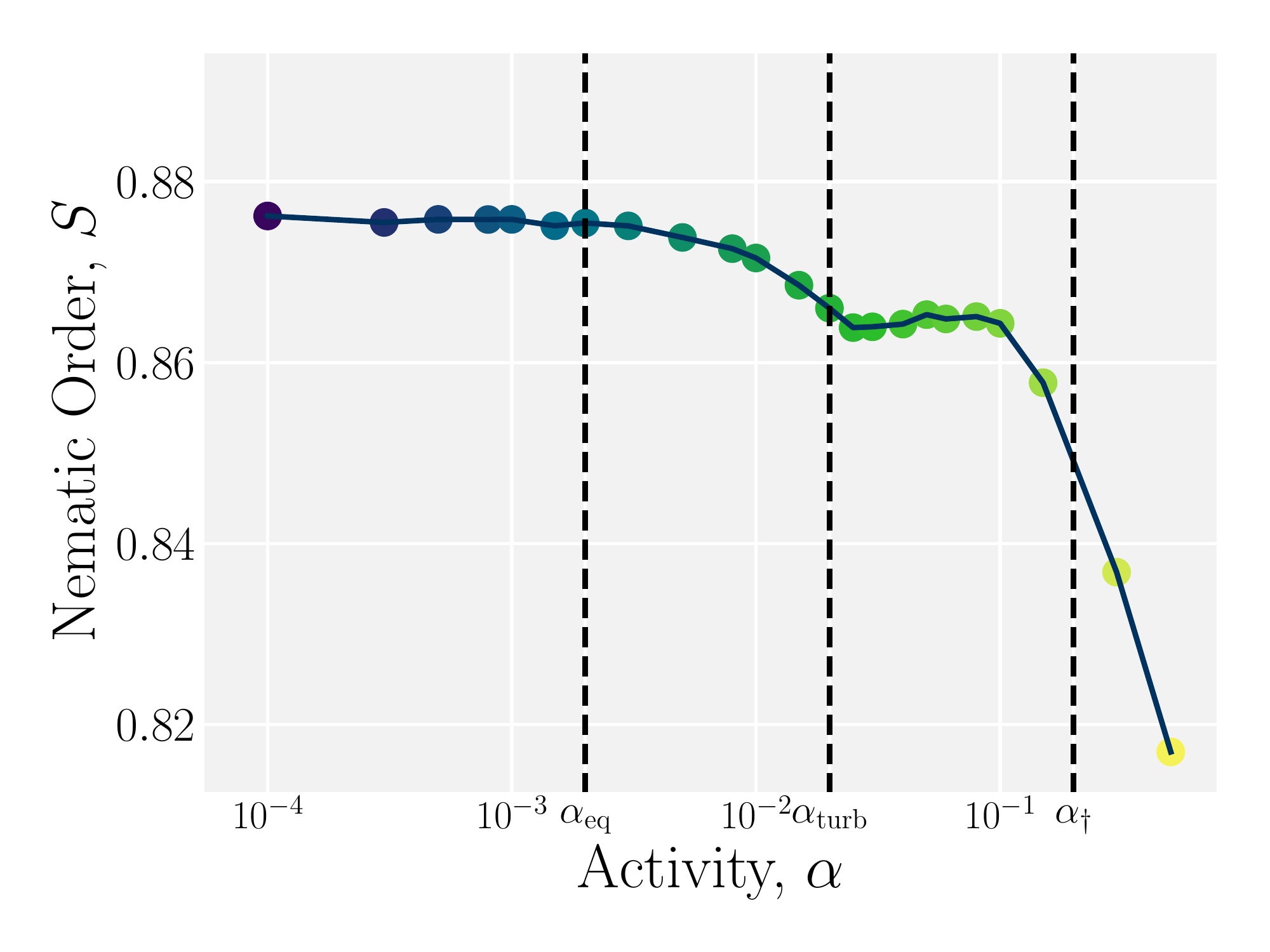}
    \end{subfigure}
    \caption{\textbf{Average nematic order in AN-MPCD} 
    Four regimes are visible, corresponding to (i) $\act \lesssim \actMIN$ nematic behaviour, (ii) $\actMIN \lesssim \act \lesssim \actMin$ onset of active effects, (iii) $\actMin \lesssim \actMax$ active turbulence, (iv) $\actMax\lesssim\act$ active turbulence scalings no longer acheived and local order plummets. 
    }
    \label{fig:nemSOrder}
\end{figure}

The distribution broadens but also changes shape with higher moments of the distribution growing more rapidly (\fig{fig:dens}a). 
Once the standard deviation begins to grow, the non-Gaussianity parameter $\chi_\text{NGM}$ also increases with activity (\fig{fig:dens}b; inset). 
This represents the widening tails of the density distributions in \fig{fig:dens}a. 
At high activities, the likelihood of finding regions with either much higher or much lower density than the mean is increased. 
This can be seen at high activities $\act \gtrsim \actMax$ with not only empty cells, but also high-density bands within a sparse nematic gas (\fig{fig:densityFluctuations} and \movie{mov:a0.0800dens}). 
As the activity is raised further, corresponding with broader density distributions, these bands appear narrower as a consequence (\movie{mov:a0.3000dens}).
The bands exhibit spatiotemporal chaos through elongation, splitting and merging, reminiscent of particle-based dry active nematic models~\cite{Chate2006PRL, PeshkovChate2014PRL, MahaultChate2019PRL}. 
However, the AN-MPCD algorithm is not observed to exhibit as substantial a decrease in nematic ordering in the gas-like phase (\fig{fig:nemSOrder}).

To better understand the variation of MPCD particles, consider the number fluctuations within sub-domains. 
From \eq{eq:GNF}, the fluctuations scale with the mean as $\sigma_{\NCell} \sim \av{\NCell}^\nu$ with $\nu=1/2$ in thermal equilibrium. 
At low activities ($\act\lesssim\actMin$), the fluctuations obey the Central Limit Theorem with $\nu=1/2$ (\fig{fig:dens}c,d). 
However, for activities in the fully developed turbulence regime, density fluctuations increase and crossover to anomalous behaviour. 
As $\nu \to 1$, giant number fluctuations dominate the statistics, indicating highly out-of-equilibrium behaviour within the active turbulence scaling regime $\actMin \lesssim \act \lesssim \actMax$ (\fig{fig:dens}d). 
The eminence of non-equilibrium particle number fluctuations in AN-MPCD highlights the mesoscale nature of the algorithm. 
AN-MPCD possesses properties of both active particle ensembles and active fluids. 

The exponent $\nu$ has a maximum at $\act = \actMax \simeq 3\times10^{-1}$. 
The decrease in $\nu$ past this point can be explained by considering how the average nematic order parameter $S$ varies with activity (\fig{fig:nemSOrder}). 
For the near-equilibrium regime of $\act \lesssim \actMIN$, the nematic order in the system is at its highest and is effectively constant. 
The system then transitions to fully-formed active turbulence regime $\actMIN \lesssim \act \lesssim \actMax$ and the nematic order transitions to another plateau at $\act \simeq \actMin$. 
At the highest activities, the nematic order drops off sharply for $\actMax \lesssim \act$, corresponding to the decrease in giant-number fluctuations, $\nu$ (\fig{fig:dens}d; \movie{mov:a0.3000dir}-\ref{mov:a0.3000vel}). 
The high activity disorders the orientation partially by causing local regions of low density, which are effectively below the isotropic-nematic transition point, leading to a drop in the nematic order. 
Conversely, the drop in nematic order causes the active nematic dipole to be more broadly distributed which disperses MPCD particles more randomly, leading to less clustering and slightly more homogeneous density structure.

\section{Conclusions}
We have proposed and quantified a mesoscopic, particle-based algorithm for simulating wet active nematics. 
It is an extension of the nematic version of the multi-particle collision dynamics (N-MPCD) method to account for active force dipoles. 
This active-dipole contribution to the MPCD collision operator injects kinetic energy but conserves translational momentum.
The active-dipole is aligned with the local nematic director. 
The strength of the force dipole is computed from the particles within each cell, resulting in a density dependent local dipole strength. 

This active-nematic momentum collision operator generates spontaneous flows, and hydrodynamic instabilities leading to defect unbinding and active turbulence. 
Activity is found to be negligible below $\actMIN$, because the Andersen thermostat absorbs the injected energy. 
In this negligible-activity limit, the fluid is indistinguishable from a passive nematic. 
For weak activity ($\actMIN\leq\act\leq\actMin$), spontaneous flows arise but the characteristic length scale of the activity is comparable to the system size and so active turbulence is not fully developed. 
In this regime, defect-pair unbinding is rare and flows correlations span the entire system. 
Only as the activity approaches $\actMin$, does the fluid begin to exhibit the aspects of fully developed active turbulence. 
In the active turbulence regime, the characteristic length scale, as measured from the defect separation, scales with activity as predicted by theory. 
Likewise, the magnitude of the fluid velocity scales with activity with an exponent comparable to the ideal prediction. 

In addition to exhibiting the traits of active turbulence as expected for a continuum model of active fluids, AN-MPCD also possesses characteristics of active particle models. 
Most prominently, the local density of MPCD particles has begun to exhibit giant number fluctuations. 
The concurrence of active nematic fluid properties and active particle properties highlights the mesoscale nature of AN-MPCD --- multi-particle collision dynamics is a coarse-grained algorithm that spans the intermediate scales between the microscopic and hydrodynamic limits. 

Much work has been accomplished studying the fundamental properties of active fluids~\cite{Marchetti2013RevModPhys-Hydrodynamics, Ramaswamy2010AnnRevCond, Bar2020AnnRev}. 
However, soft condensed matter physics is often not principally interested in the dynamics of a background solvent itself. 
Rather, relaxational dynamics, self-assembly, transport of complex solutes embedded within a fluidic medium, or other multiscale phenomenta are the subjects of physical interest. 
Active-Nematic MPCD opens the door to simulating complex particles embedded within a spontaneously flowing active medium. 
Future work could simulate of Janus particles~\cite{Loewe2021NJP} or other anisotropic passive colloids, including flexible filaments or even polymeric materials~\cite{Gompper2022PRE}. 
Similarly studies of driven-particles within active nematics~\cite{MarenduzzoPRL2012, Leheny2020SoftMatter} could be extended. 
In fact, AN-MPCD could be used to study suspensions of active particles, such as swimming bacteria, suspended in an active solvent with a different activity or symmetry --- \eg pushers in contractile nematics. 
By bridging microscopic models, such as active Brownian particles, and macroscopic models, such as the Toner-Tu equation, AN-MPCD offers a pathway for numerical studies of novel active hybrid materials. 

\vspace{\baselineskip}
\begin{acknowledgments}
We thank Julia Yeomans for useful discussions and Louise Head for suggestions on algorithm development. 
This research has received funding from the European Research Council (ERC) under the European Union’s Horizon 2020 research and innovation programme (Grant agreement No. 851196) and EMBO (ALTF181-2013). 
\end{acknowledgments}

\appendix
\section{Movie Captions}
\begin{enumerate}
    \item \label{mov:defUnbindDir} 
    Director field coloured by nematic order parameter $S_c$ demonstrating defect unbinding.
    $\act=0.03$, $\lenSys=30$, simulation length is $1000\delta t$, and each frame is $5\delta t$ apart.
    \item \label{mov:a0.0008dir} 
    Director field coloured by nematic order parameter $S_c$ for $\act < \actMIN$; activity value $\act=0.0008$. 
    The system size is $\lenSys=100$ and duration $2500\delta t$ following a $2000 \delta t$ warmup. Each frame is $5\delta t$ apart.
    \item \label{mov:a0.0080dir} 
    Director field coloured by nematic order parameter $S_c$ for $\actMIN < \act < \actMin$; activity value $\act=0.008$. 
    All other parameters are the same as \movie{mov:a0.0008dir}.
    \item \label{mov:a0.0800dir} 
    Director field coloured by nematic order parameter $S_c$ for $\actMin < \act < \actMax$; activity value $\act=0.08$.
    All other parameters are the same as \movie{mov:a0.0008dir}.
    \item \label{mov:a0.0008vel}
    Velocity field coloured by speed $\abs{\vec{v}}$ corresponding to \movie{mov:a0.0008dir} with activity $\act=0.0008$.
    \item \label{mov:a0.0080vel}
    Velocity field coloured by speed $\abs{\vec{v}}$ corresponding to \movie{mov:a0.0080dir} with activity $\act=0.008$.
    \item \label{mov:a0.0800vel}
    Velocity field coloured by speed $\abs{\vec{v}}$ corresponding to \movie{mov:a0.0800dir} with activity $\act=0.08$.
    \item \label{mov:a0.0800dens} 
    Number density field $N_C/N_C^\mathrm{max}$, where $N_C^\mathrm{max}$ is the instantaneous maximum cell population.
    The simulation in this movie is the same as \movie{mov:a0.0800dir} for $\act=0.08$, with identical parameters and frame times.
    \item \label{mov:a0.3000dens} 
    Number density field $N_C/N_C^\mathrm{max}$ for $\actMax < \act$; activity value $\act=0.3$.
    All other parameters are the same as \movie{mov:a0.0800dens}
    \item \label{mov:a0.3000dir} 
    Director field coloured by nematic order parameter $S_c$ corresponding to \movie{mov:a0.3000dens} with activity $\act=0.3$. 
    \item \label{mov:a0.3000vel}
    Velocity field coloured by speed  $\abs{\vec{v}}$ corresponding to \movie{mov:a0.3000dens} with activity $\act=0.3$.
\end{enumerate}

\bibliography{biblio.bib}

\end{document}